\documentclass[]{article}
\usepackage[T1]{fontenc}
\usepackage[latin9]{inputenc}
\usepackage{geometry}
\usepackage{amsmath}
\usepackage{amssymb}
\usepackage{graphicx}
\usepackage{braket}
\usepackage{color}
\usepackage{placeins}

\makeatletter

\providecommand{\tabularnewline}{\\}

\numberwithin{equation}{section}
\numberwithin{figure}{section}
\newtheorem{thm}{\protect\theoremname}
  \newtheorem{rem}[thm]{\protect\remarkname}

\makeatother

\usepackage[english]{babel}
  \providecommand{\remarkname}{Remark}
\providecommand{\theoremname}{Theorem}

\begin{document}

\global\long\def\ve{\varepsilon}
\global\long\def\R{\mathbb{R}}
\global\long\def\Rn{\mathbb{R}^{n}}
\global\long\def\Rd{\mathbb{R}^{d}}
\global\long\def\E{\mathbb{E}}
\global\long\def\P{\mathbb{P}}
\global\long\def\bx{\mathbf{x}}
\global\long\def\vp{\varphi}
\global\long\def\ra{\rightarrow}
\global\long\def\smooth{C^{\infty}}
\global\long\def\Tr{\mathrm{Tr}}
\global\long\def\u{\uparrow}
\global\long\def\d{\downarrow}

\newcommand{\bvec}[1]{\mathbf{#1}}
\renewcommand{\Re}{\mathrm{Re}}
\renewcommand{\Im}{\mathrm{Im}}

\newcommand{\mc}[1]{\mathcal{#1}}
\newcommand{\mf}[1]{\mathfrak{#1}}

\newcommand{\vr}{\bvec{r}}
\newcommand{\vF}{\bvec{F}}
\newcommand{\vg}{\bvec{g}}
\newcommand{\vR}{\bvec{R}}
\newcommand{\vq}{\bvec{q}}
\newcommand{\vx}{\bvec{x}}
\newcommand{\ud}{\,\mathrm{d}}
\newcommand{\abs}[1]{\lvert#1\rvert}
\newcommand{\norm}[1]{\lVert#1\rVert}
\newcommand{\wt}[1]{\widetilde{#1}}

\newcommand{\Or}{\mathcal{O}}
\newcommand{\mcF}{\mathcal{F}}
\newcommand{\I}{\mathrm{i}}
\newcommand{\EE}{\mathbb{E}}
\newcommand{\NN}{\mathbb{N}}
\newcommand{\RR}{\mathbb{R}}
\newcommand{\CC}{\mathbb{C}}
\newcommand{\ZZ}{\mathbb{Z}}

\newcommand{\LL}[1]{\textcolor{red}{[LL:#1]}}
\newcommand{\ML}[1]{\textcolor{blue}{[ML:#1]}}

\title{Variational embedding for quantum many-body problems}

\author{
Lin Lin\thanks{Department of Mathematics, University of California, Berkeley, Berkeley, CA 94720 and Computational Research Division, Lawrence Berkeley National Laboratory, Berkeley, CA 94720. Email: \texttt{linlin@math.berkeley.edu}}
\and Michael Lindsey\thanks{Department of Mathematics, Courant Institute of Mathematical Sciences, New York University, New York, NY 10012. Email: \texttt{michael.lindsey@cims.nyu.edu}}
}
\maketitle

\begin{abstract}
Quantum embedding theories are powerful tools for approximately solving large-scale strongly correlated quantum many-body problems. The main idea of quantum embedding is to glue together a highly accurate quantum theory at the local scale and a less accurate  quantum theory at the global scale. We introduce the first quantum embedding theory that is also variational, in that it is guaranteed to provide a one-sided bound for the exact ground-state energy.  Our method, which we call the variational embedding method, provides a lower bound for this quantity. The method relaxes the representability conditions for quantum marginals to a set of linear and semidefinite constraints that operate at both local and global scales, resulting in a semidefinite program (SDP) to be solved numerically. The accuracy of the method can be systematically improved. The method is versatile and can be applied, in particular, to quantum many-body problems for both quantum spin systems and fermionic systems, such as those arising from electronic structure calculations. We describe how the proper notion of quantum marginal, sufficiently general to accommodate both of these settings, should be phrased in terms of certain algebras of operators. 
We also investigate the duality theory for our SDPs, which offers valuable perspective on our method as an embedding theory. As a byproduct of this investigation, we describe a formulation for efficiently implementing the variational embedding method via a partial dualization procedure and the solution of quantum analogs of the Kantorovich problem from optimal transport theory.
\end{abstract}


\section{Introduction}

Quantum many-body problems, such as the problem of computing the 
ground state of a system of quantum spins or fermions, have far-reaching applications in
physics, chemistry, materials science, and beyond. Certain such 
problems, including those involving fermions in the `strongly correlated' regime, are among the most
challenging problems in scientific computing. Roughly speaking, a ground state of 
a quantum many-body problem is specified by
a wavefunction $\ket{\Phi}$ obtained as a minimizer
of the following optimization problem:
\begin{equation}
  E_{0}=\min_{\ket{\Phi}\in\mc{H},\,\braket{\Phi|\Phi}=1}\braket{\Phi|\hat{H}|\Phi},
  \label{eqn:minProblem}
\end{equation}
in which we have employed the Dirac bra-ket notation, and $\mc{H}$  is the Hilbert space 
whose elements are quantum states. 
The optimization problem~\eqref{eqn:minProblem} is 
equivalent to a linear eigenvalue problem, with the 
ground state $\ket{\Phi}$ given by the
eigenvector corresponding to the smallest eigenvalue (assuming the eigenvalue is simple) of $\hat{H}$.  The
cost of directly finding $\ket{\Phi}$ 
generally scales exponentially with respect to the system size. It is therefore of
paramount interest to reduce the computational complexity of this task by accepting some
controlled sacrifice of accuracy.

Among all the approaches to solving the problem~\eqref{eqn:minProblem}, some are 
\emph{variational} in the sense that they provide an approximation for $E_0$ which is guaranteed to be 
either an upper or lower bound. For example, methods which restrict the optimization over 
$\vert\Phi \rangle \in \mc{H}$ to some computationally tractable subset provide 
\emph{upper bounds} for $E_0$. Examples of such methods include 
the Hartree-Fock
approximation~\cite{SzaboOstlund1989}, matrix product states (MPS) (also known as
tensor trains)~\cite{White1992,OseledetsTyrtyshnikov2009}, and other tensor network methods such as projected entangled-pair states (PEPS)~\cite{VerstraeteCirac2004,Orus2014}. Meanwhile, other approaches attempt to formulate  
tractable \emph{relaxations} of the variational principle~\eqref{eqn:minProblem}. The 
idea of such approaches is to reformulate~\eqref{eqn:minProblem} as an equivalent optimization 
problem in terms of density matrices, in which the difficulty is encoded in the constraints, and then to enforce only a computationally
tractable subset of these constraints. Such procedures yield guaranteed \emph{lower bounds} for $E_0$.
The most well-known example of such an approach is the two-electron reduced density matrix (2-RDM) theory for fermionic systems~\cite{Mazziotti1998,Mazziotti2004,CancesStoltzLewin2006,Mazziotti2012,ZhaoBraamsFukudaEtAl2004,LiWenYangEtAl2018,AndersonNakataIgarashiEtAl2013,NakataNakatsujiEharaEtAl2001,DeprinceMazziotti2010}. 

Another category of approaches to the quantum many-body problem is that of the 
\emph{quantum embedding theories}~\cite{SunChan2016}. Notable examples
include the dynamical mean-field theory (DMFT)~\cite{GeorgesKotliarKrauthEtAl1996,KotliarSavrasovHauleEtAl2006}
and the 
density matrix embedding theory (DMET)~\cite{KniziaChan2012,KniziaChan2013}. These methods divide the
global system into a set of local clusters (sometimes called
fragments), where the size of each cluster is taken to be independent of the global system size. Then one
derives a modified quantum many-body problem for each cluster, which can
be solved directly or using approximate (but highly accurate) methods. 
The information from all of the clusters is then
`glued' together using global reduced quantities, such as the one-electron reduced density matrix (1-RDM) in DMET, or the 
single-particle Green's function in DMFT. The method can
be solved self-consistently to remove the discrepancy between these 
global quantities and local fragment data.

In this work we propose an approach to the quantum many-body problem, which is the first example 
to our knowledge of a quantum embedding method that is also variational. We 
therefore call it the variational embedding method, which we 
develop below for quantum spin systems and second-quantized fermionic systems. 
(Note that our framework for quantum spin systems formally includes the setting 
of second-quantized bosonic systems as an infinite-dimensional limit.)
The fundamental objects considered in our approach are quantum marginals, 
which are defined with respect 
to a decomposition of the global system into clusters. 
The quantum marginals are referred to as such 
because they are analogous to marginal distributions in the setting of classical probability theory. 
In the setting of quantum spins, these are just the reduced density operators, which 
are defined as partial traces of a global density operator. In the fermionic setting, 
a more general perspective
is introduced to define the analogous quantities. This perspective
views marginals as functionals on appropriate operator algebras.

Our approach is in particular a relaxation of the variational principle~\eqref{eqn:minProblem}, hence 
yields a \emph{lower bound} for $E_0$. It is an embedding method in the sense that clusters are represented 
with high fidelity and glued together via some reduced global data. 
The accuracy of the variational embedding method can be
systematically improved 
by increasing the cluster size, \emph{or} by considering marginals for larger groups
of clusters, e.g., pairs, triples, etc. The relaxed optimization problem defining
the variational embedding method is a semidefinite program (SDP), whose cost
scales polynomially with respect to the system size (for fixed cluster size). Treating this relaxation 
as the primal problem, we derive the dual problem and show that
the duality gap is zero. We also introduce a partial dualization of the primal problem, in which 
the interpretation as an embedding method becomes even clearer. In particular, we 
see the emergence of effective Hamiltonians for embedded problems, which are coupled 
only via the global determination of these effective Hamiltonians. The embedded problems 
are themselves quantum analogs of the Kantorovich problem of optimal transport~\cite{villani2008optimal}. Although 
our presentation of this quantum Kantorovich problem, which emphasizes general cost operators,
differs somewhat from that of the existing literature, the same basic problem has appeared 
in~\cite{ChenGangboEtAl2019,GolseEtAl2016,DattaRouze2017,ZhouYingEtAl2019,CagliotiGolsePaul2018}.

We also describe how variational embedding adapts to the scenario of
overlapping clusters.  It can be seen readily that allowing for overlapping
clusters tightens the constraints, yielding tighter lower bounds for the ground-state energy at comparable computational cost.  This point
may be of interest because the value-add of overlapping clusters in embedding theories such as DMET and DMFT is not yet clear~\cite{BiroliParcolletKotliar2004,YeRickeTranEtAl2019}.  We also describe how translation invariance can be exploited 
in the implementation of variational embedding.

As proof-of-principle, we demonstrate the performance of the
variational embedding method for two quantum spin models (the transverse Ising model 
and the
anti-ferromagnetic Heisenberg model) and one fermionic model (the Hubbard model). The system
size is small due to the limitations of the preliminary implementation in
\textsf{CVX}~\cite{grant2008cvx} within \textsf{MATLAB}, and we plan to develop more efficient implementations to 
accommodate larger systems in the near future. 
In the numerical 
experiments, we solve the primal problem directly, but the partial dualization mentioned above 
suggests more efficient methods for solving the variational
embedding method, with tractable scaling for extended systems.

\subsection{Related work}
In the fermionic setting, the aforementioned 2-RDM theory is the closest relative of variational embedding. Nonetheless, we point out that our `fermionic marginals' are different from the 2-RDM. In general, neither the variational embedding method nor the 2-RDM theory adopts a strictly tighter relaxation than the other. Roughly speaking, the variational embedding method enforces tighter constraints `within clusters' but weaker constraints `across clusters,' relative to the most accurate 2-RDM theories. Therefore we expect that variational embedding can be more efficient for treating strong correlation effects that are relatively local in nature.  That said, both frameworks are highly modular.  In fact, it may be possible to adapt existing 2-RDM theories as methods for solving the embedded problems obtained in the variational embedding method. Finally, we comment that the partial dual formulation holds promise for scaling to extended systems, where 2-RDM theories can become prohibitively computationally expensive. 

The approach of this paper can also be understood as an approximate method for 
solving the `quantum marginal problem,'~\cite{Klyachko2006,Schilling2013} i.e., the problem of determining whether a set of 
quantum marginals could have been obtained from a global quantum density operator. 
In general, the exact solution of this problem is intractable, so approximate methods must be adopted.

Finding approximate solutions to the quantum marginal problem can be viewed as a 
quantum analog of the problem of finding outer bounds to the marginal polytope in classical 
probability~\cite{WainwrightJordan}. 
In our approach, we derive two main types of constraints: local consistency constraints (which are linear) and global
semidefinite constraints. The local consistency constraints, which enforce compatibility 
between marginals that share sites, are so termed by 
analogy to the constraints of the same name appearing in relaxations of the 
classical marginal polytope~\cite{WainwrightJordan}. These constraints alone can be viewed 
as underlying the belief propagation (BP)~\cite{Pearl1982} approximation for classical graphical models (see, also, 
e.g.,~\cite{WainwrightJordan} for reference). Note with caution that BP should be thought of as 
an \emph{algorithm}, in addition to a set of modeling assumptions. Also note that BP involves an 
implicit approximation of the entropy, which is not relevant in the zero-temperature setting, i.e., the 
setting of this work.

BP has been generalized to the quantum setting (specifically, the setting of quantum spin systems in the sense of this paper)~\cite{LeiferPoulin2008}, and 
other works~\cite{PoulinHastings2011,FerrisPoulin2013} have more carefully studied quantum entropy approximation for 
quantum spin systems in the context of the local consistency constraints that are featured in BP. 
Meanwhile,~\cite{BarthelHubener2012} considers a semidefinite relaxation in a zero-temperature, translation-invariant setting for both quantum spins and fermions. In our language, one can view~\cite{BarthelHubener2012} as implicitly considering overlapping clusters for which local consistency constraints (which are generally more complicated to enforce due to cluster overlap) are automatically satisfied without need for explicit enforcement due to the translation invariance. None of these cluster-based works can be viewed as considering an analog of the global semidefinite constraints introduced in this work. Moreover, 
these works only support local Hamiltonians and cannot support long-range (e.g., Coulomb-type) interactions. 
In fact, the global semidefinite constraints improve the quality of the relaxation even in the case of local Hamiltonians (as we shall 
demonstrate in Section~\ref{sec:numerics} below), but more dramatically they open the door to 
 cluster-based semidefinite relaxations for long-range Hamiltonians and potentially \emph{ab initio} electronic structure 
 problems.

Another point of comparison is the 
Lasserre hierarchy~\cite{lasserreBook,WainwrightJordan} of semidefinite relaxations, often considered as means for approximating the marginal polytope in classical probability. Our method is not the quantum analog of any relaxation from this Lasserre hierarchy in the classical setting, nor is our method recovered from the Lasserre hierarchy as applied directly to the quantum many-body problem. In fact, the variational embedding method can be understood as advancing different 
systematically improvable hierarchies, both in the cluster size and in the sizes of the groups of clusters for which marginals are considered.

The variational embedding method can also be understood as a way to tighten the variational lower bound obtained in~\cite{KhooEtAl2019} 
for fermionic many-body problems based on the strictly correlated electron (SCE) formulation \cite{SeidlPerdewLevy1999,SeidlGori-GiorgiSavin2007}.  There are two sources of error in the approach of~\cite{KhooEtAl2019}: a model error (which only vanishes in the `strictly correlated' limit of infinitely strong Coulomb repulsion) and an additional relaxation error 
that emerges from the relaxation of a \emph{classical} marginal problem.
The variational embedding method introduced in this paper can be viewed as a 
fully quantum version of this relaxation. It
avoids any analogous notion of model error and 
can be shown to provide energies at least as tight as those obtained in~\cite{KhooEtAl2019}. 

\subsection{Outline}
In section~\ref{sec:quantumSpins} we formulate variational embedding for quantum spin systems. 
After preliminary discussion in section~\ref{sec:spinPrelim}, we go on to introduce the local consistency 
constraints and global semidefinite constraints in sections ~\ref{sec:localConsistency} and 
\ref{sec:globalSemidefinite}, respectively. In section~\ref{sec:abstractGlobal} we discuss a 
more abstract perspective on the global semidefinite constraints that is, in particular, 
more portable to the fermionic setting to appear later on. In section~\ref{sec:higherMarginal} we introduce 
variational embedding constraints for higher marginals (i.e., marginals for higher tuples of 
sites), and in section~\ref{sec:cluster} we introduce the cluster perspective on variational embedding. 
In section~\ref{sec:overlapping} we discuss how variational embedding can accommodate overlapping 
clusters for tighter relaxations, and in section~\ref{sec:translationInvariant} we discuss how 
translation-invariance can be exploited, as well as additional `periodicity constraints' that can be 
imposed in this setting.

Section~\ref{sec:fermions} concerns the formulation of variational embedding for fermionic 
systems in second quantization. After discussing preliminaries in section~\ref{sec:fermionPrelim}, 
we employ the language of star-algebras to define appropriate 
fermionic marginals in section~\ref{sec:abstractFerm}. Using this language, we provide an 
abstract formulation of variational embedding for fermions in section~\ref{sec:ferm2mar}, which 
we show is exact for non-interacting problems (i.e., problems specified by single-body Hamiltonians) 
in section~\ref{sec:nonintExact}. In section~\ref{sec:concrete}, we demonstrate how the abstract formulation can be practically implemented as a SDP.

Section~\ref{sec:numerics} presents various numerical experiments. In sections~\ref{sec:tfi}, 
\ref{sec:afh}, and \ref{sec:hubbard} we treat the transverse-field Ising, anti-ferromagnetic Heisenberg, and 
Hubbard models, respectively.

Finally, we conclude in section~\ref{sec:duality} with a discussion of duality for the SDP of
variational embedding. To prepare for the formulation of the dual problem, we discuss 
in section~\ref{sec:kantorovich}
a quantum analog of the Kantorovich problem from optimal transport. Then in section~\ref{sec:partialDual} 
we introduce a partially dualized SDP, which reveals that the variational embedding solution 
can be obtained as the solution of several quantum Kantorovich problems specified by 
`effective Hamiltonians,' which are completely decoupled from one another apart from the determination 
of these effective Hamiltonians. In section~\ref{sec:compPartialDual} we discuss the computational 
implications of this observation, and in section~\ref{sec:fullDual} we close with a derivation of the 
full dual problem and a discussion of strong duality.

\subsection*{Acknowledgments} 

This work was partially supported by the Department of Energy under Grant No. DE-SC0017867, No. DE-AC02-05CH11231 (L.L.), by the Air Force Office of Scientific Research under award number FA9550-18-1-0095 (L.L. and M.L.), by the National Science Foundation Graduate Research Fellowship Program under grant DGE-1106400 and the National Science Foundation under Award No. 1903031 (M.L.).  We thank Garnet Chan, Jianfeng Lu and Lexing Ying for helpful discussions.

\section{Quantum spins}

\label{sec:quantumSpins}

\subsection{Preliminaries}
\label{sec:spinPrelim}
Let $i=1,\ldots,M$ index the sites, and for each site
$i$ let $X_{i}$ be the classical state space (discrete, for simplicity).
For each site, the quantum state space is $Q_{i}:=\mathbb{C}^{X_{i}}$,
and the global quantum state space is 
\[
\mathcal{Q}:=\bigotimes_{i=1}^{M}Q_{i}\simeq\mathbb{C}^{\mathcal{X}},
\]
 where $\mathcal{X}:=\prod_{i=1}^{M}X_{i}$. Let $H_{i}$ denote a
Hermitian operator $Q_{i}\ra Q_{i}$, and let $H_{ij}$ denote a Hermitian
operator $Q_{i}\otimes Q_{j}\ra Q_{i}\otimes Q_{j}$. We will use
the hatted notation $\hat{H}_{i}$ to denote the operator $\mathcal{Q}\ra\mathcal{Q}$
obtained by tensoring $H_{i}$ by the identity operator on all sites
$k\neq i$, and likewise we identify $\hat{H}_{ij}$ with the operator
$\mathcal{Q}\ra\mathcal{Q}$ obtained by tensoring $H_{ij}$ with
the identity on all sites $k\notin\{i,j\}$. Then we consider a Hamiltonian
$\hat{H}:\mathcal{Q}\ra\mathcal{Q}$ of the form 
\[
\hat{H}=\sum_{i}\hat{H}_{i}+\sum_{i<j}\hat{H}_{ij}.
\]
 
\begin{rem}
We shall introduce several examples of interest in the case $X_{i}=\{-1,1\}$,
i.e., the case of quantum spin-$\tfrac{1}{2}$ systems. 
The Pauli matrices
\[
\sigma^{x}=\left(\begin{array}{cc}
0 & 1\\
1 & 0
\end{array}\right),\quad\sigma^{y}=\left(\begin{array}{cc}
0 & -i\\
i & 0
\end{array}\right),\quad\sigma^{x}=\left(\begin{array}{cc}
1 & 0\\
0 & -1
\end{array}\right),
\]
together with the identity $I_{2}$, form a basis for Hermitian
operators on $\mathbb{C}^{2}$. Now let $\sigma_{i}^{x/y/z}\in\mathcal{H}(\bigotimes_{i}\mathbb{C}^{2})\simeq\bigotimes_{i}\mathcal{H}(\mathbb{C}^{2})$ be obtained by tensoring a copy of $\sigma^{x/y/z}$ for the $i$-th site with the identity $I_{2}$ on all the other sites. 
Two examples of the quantum spin systems are the transverse-field Ising (TFI)
Hamiltonian and anti-ferromagnetic Heisenberg (AFH) Hamiltonian, specified by the
Hamiltonians 
\begin{equation}
\hat{H}_{\mathrm{TFI}}=-h\sum_{i}\sigma_{i}^{x}-\sum_{i \sim j}\sigma_{i}^{z}\sigma_{j}^{z},\label{eq:tfi}
\end{equation}
\begin{equation}
\hat{H}_{\mathrm{AFH}}=\sum_{i\sim j}\left[\sigma_{i}^{x}\sigma_{j}^{x}+\sigma_{i}^{y}\sigma_{j}^{y}+\sigma_{i}^{z}\sigma_{j}^{z}\right].\label{eq:afh}
\end{equation}
where the summation of $i \sim j$ indicates summation over all pairs of indices that 
are adjacent in a graph defined on the index set (usually the graph is a square lattice).
In the TFI Hamiltonian, $h\in\R$ is a scalar parameter.
\end{rem}

We are interested in computing the ground-state energy 
\[
E_{0}=\inf\left\{ \langle\Phi\vert\hat{H}\vert\Phi\rangle\,:\,\vert\Phi\rangle\in\mathcal{Q},\,\langle\Phi\vert\Phi\rangle=1\right\}.
\]
It can be equivalently recast as 
\[
E_{0}=\inf_{\rho\in\mathcal{D}(\mathcal{Q})}\Tr[\hat{H}\rho],
\]
where $\mathcal{D}(\mathcal{Q})$ denotes the set of density operators
on $\mathcal{Q}$ (i.e., positive semidefinite linear operators $\mathcal{Q}\ra\mathcal{Q}$
of unit trace). Assuming that there exists a unique ground state $\vert\Phi_{0}\rangle$,
the infimum is attained at $\rho=\vert\Phi_{0}\rangle\langle\Phi_{0}\vert$.
Now we can write 
\begin{equation}
E_{0}=\inf_{\{\rho_{ij}\}_{i<j}\in\mathbf{QM}_{2}(\mathcal{Q})}\left(\sum_{i}\Tr\left[H_{i}\rho_{i}\right]+\sum_{i<j}\Tr\left[H_{ij}\rho_{ij}\right]\right),\label{eq:minRep}
\end{equation}
 where $\mathbf{QM}_{2}(\mathcal{Q})$ denotes the set of collections
$\{\rho_{ij}\}_{i<j}$ of representable quantum two-marginals, i.e.,
those collections $\{\rho_{ij}\}$ which can be obtained as reduced
density operators of a single $\rho\in\mathcal{D}(\mathcal{Q})$ via
the partial trace, as in 
\[
\rho_{ij}=\Tr_{\{1,\ldots,M\}\backslash\{i,j\}}[\rho],
\]
 where $i<j$.\\

To clarify, here we view $\rho$ as being equipped with labels $1,\ldots,M$
for its indices as $\rho=\rho_{i_{1}\cdots i_{M},j_{1}\cdots j_{M}}$,
and for any subset $S\subset\{1,\ldots,M\}$, $\rho_{S}=\Tr_{\{1,\ldots,M\}\backslash S}[\rho]$
denotes the reduced density operator obtained by tracing out the indices
contained in $S$, with the remaining labels maintained. We comment
that the partial trace $\rho_{S}$ may be equivalently defined as
the unique operator on $\bigotimes_{i\in S}Q_{i}$ such that $\Tr[\hat{A}\rho_{S}]=\Tr[\hat{A}\rho]$
for all operators $\hat{A}$ on $\bigotimes_{i\in S}Q_{i}$ (alternatively
viewed as operators on $\mathcal{Q}$ by tensoring with the identity).
This perspective illustrates the relationship between marginalization
in the quantum spin setting (i.e., computing the partial trace) and
the more abstract notion of marginalization that is necessary for
the treatment of fermions in section \ref{sec:fermions} below.\\

For convenience, we denote $\rho_{ij}=\rho_{\{i,j\}}$ for $i<j$
as above. It is convenient to then define $\rho_{ij}$ for $i>j$
via the stipulation that $\sigma_{ij}\rho_{ij}\sigma_{ji}=\rho_{ji}$,
where $\sigma_{ij}:Q_{i}\otimes Q_{j}\ra Q_{j}\otimes Q_{i}$ is the
linear operator defined by $\sigma_{ij}(\phi_{i}\otimes\phi_{j})=\phi_{j}\otimes\phi_{i}$.
Finally, we remark that the one-marginals $\rho_{i}=\Tr_{\{1,\ldots,M\}\backslash\{i\}}[\rho]$
are determined by the two-marginals via $\rho_{i}=\Tr_{\{j\}}[\rho_{ij}]$,
and this dependence is meant to be understood implicitly in (\ref{eq:minRep}).
We will occasionally denote $\rho_{ii}:=\rho_{i}$.

\subsection{Local consistency constraints}
\label{sec:localConsistency}
Now it is of interest to determine necessary conditions satisfied
by collections in $\mathbf{QM}_{2}(\mathcal{Q})$. By enforcing a
set of necessary conditions as a proxy for membership in $\mathbf{QM}_{2}(\mathcal{Q})$,
we can obtain a lower bound on the ground state energy.\\

To begin with, the $\rho_{ij}$ are themselves density operators on
$Q_{i}\otimes Q_{j}$, i.e., $\rho_{ij}\succeq0$ with $\Tr[\rho_{ij}]=1$.
Moreover, we must have $\Tr_{j}[\rho_{ij}]=\Tr_{j'}[\rho_{ij'}]$
for all $i$ and $j,j'\neq i$, and we must have $\sigma_{ij}\rho_{ij}\sigma_{ji}=\rho_{ji}$.
These constraints define the set of \emph{locally consistent quantum
two-marginals}. Call this set $\mathbf{LQM}_{2}(\mathcal{Q})$. In
practice we define auxiliary variable $\rho_{i}$ for the one-marginals,
constrained to satisfy $\rho_{i}=\Tr_{j}[\rho_{ij}]=\Tr_{i}[\rho_{ji}]$.
The constraints $\Tr[\rho_{ij}]=1$ for all $i,j$ can in fact be
enforced by requiring $\Tr[\rho_{i}]=1$ for all $i$, since $\Tr[\rho_{ij}]=\Tr[\Tr_{j}[\rho_{ij}]]$.\\

Note that the local consistency constraint $\Tr_{j}[\rho_{ij}]=\rho_{i}$
is equivalent to insisting that $\Tr[\hat{A}\rho_{ij}]=\Tr[\hat{A}\rho_{i}]$
for all operators $\hat{A}$ on $Q_{i}$ (considered also as operators
on $Q_{i}\otimes Q_{j}$ by tensoring with the identity). This perspective
highlights the connection to the abstract local consistency constraints
appearing in the discussion of fermionic systems in section \ref{sec:fermions}
below.

\subsection{Global semidefinite constraints and the two-marginal SDP}
\label{sec:globalSemidefinite}
We can derive a further constraint, more global in nature, as follows.
Consider operators $\hat{O}:\mathcal{Q}\ra\mathcal{Q}$ (not necessarily
Hermitian) of the form $\hat{O}=\sum_{i}\hat{O}_{i}$, where each
$\hat{O}_{i}$ is a one-body operator on $\mathcal{Q}$, i.e., obtained
by tensoring an operator $O_{i}$ on $Q_{i}$ with the identity. Now
$\hat{O}^{\dagger}\hat{O}\succeq0$, so 
\begin{equation}
\Tr\left[\rho\,\hat{O}^{\dagger}\hat{O}\right]\geq0\label{eq:sum1bIneq}
\end{equation}
 for any $\rho\in\mathcal{D}(\mathcal{Q})$. We will expand the left-hand
side to obtain a constraint on the quantum two-marginals, which can
be phrased as a semidefinite matrix constraint. First compute
\begin{eqnarray*}
0 & \leq & \Tr\left[\rho\,\hat{O}^{\dagger}\hat{O}\right]\\
 & = & \Tr\left[\rho\sum_{ij}\hat{O}_{i}^{\dagger}\hat{O}_{j}\right]\\
 & = & \sum_{i}\Tr\left[\rho_{i}O_{i}^{\dagger}O_{i}\right]+\sum_{i\neq j}\Tr\left[\rho_{ij}O_{i}^{\dagger}\otimes O_{j}\right].
\end{eqnarray*}
 Now without loss of generality, we can identify $X_{i}$ with $\{1,\ldots,m_{i}\}$
where $m_{i}=\vert X_{i}\vert$. Hence we can think of $O_{i}$ as
an arbitrary complex matrix $O_{i}=\left(O_{i,kl}\right)_{k,l=1,\ldots,m_{i}}$.
We will use square brackets to indicate entries of an operator as
in $[O_{i}]_{kl}=O_{i,kl}$. Note that the two-marginal $\rho_{ij}$
is an operator $Q_{i}\otimes Q_{j}\ra Q_{i}\otimes Q_{j}$, so we
denote its $((k,p),(l,q))$ entry by $[\rho_{ij}]_{kp,lq}$ for $k,l=1,\ldots,m_{i}$
and $p,q=1,\ldots,m_{j}$. Finally, for $i\neq j$, observe that 
\begin{eqnarray*}
\left[O_{i}^{\dagger}\otimes O_{j}\right]_{kp,lq} & = & [O_{i}^{\dagger}]_{kl}[O_{j}]_{pq}\\
& = & \overline{O_{i,lk}}O_{j,pq}.
\end{eqnarray*}
Then we expand the $i\neq j$ sum to obtain 
\begin{eqnarray*}
\sum_{i\neq j}\Tr\left[\rho_{ij}\,O_{i}^{\dagger}\otimes O_{j}\right] & = & \sum_{i\neq j}\sum_{k,l=1}^{m_{i}}\sum_{p,q=1}^{m_{j}}[\rho_{ij}]_{lq,kp}\left[O_{i}^{\dagger}\otimes O_{j}\right]_{kp,lq}\\
 & = & \sum_{i\neq j}\sum_{k,l=1}^{m_{i}}\sum_{p,q=1}^{m_{j}}[\rho_{ij}]_{lq,kp}\overline{O_{i,lk}}O_{j,pq}\\
 & = & \sum_{i,j=1}^{M}\sum_{k,l=1}^{m_{i}}\sum_{p,q=1}^{m_{j}}(1-\delta_{ij})[\rho_{ij}]_{lq,kp}\overline{O_{i,lk}}O_{j,pq}.
\end{eqnarray*}
 Next expand the $i$ sum: 
\begin{eqnarray*}
\sum_{i}\Tr\left[\rho_{i}\,O_{i}^{\dagger}O_{i}\right] & = & \sum_{i}\sum_{k=1}^{m_{i}}\sum_{q=1}^{m_{i}}[\rho_{i}]_{qk}\left[O_{i}^{\dagger}O_{i}\right]_{kq}\\
 & = & \sum_{i}\sum_{k,l=1}^{m_{i}}\sum_{q=1}^{m_{i}}[\rho_{i}]_{qk}[\hat{O}_{i}^{\dagger}]_{kl}[\hat{O}_{i}]_{lq}\\
 & = & \sum_{i}\sum_{k,l=1}^{m_{i}}\sum_{q=1}^{m_{i}}[\rho_{i}]_{qk}\overline{O_{i,lk}}O_{i,lq}\\
 & = & \sum_{i}\sum_{k,l=1}^{m_{i}}\sum_{p,q=1}^{m_{i}}\delta_{lp}[\rho_{i}]_{qk}\overline{O_{i,lk}}O_{i,pq}\\
 & = & \sum_{i,j=1}^{M}\sum_{k,l=1}^{m_{i}}\sum_{p,q=1}^{m_{j}}\delta_{ij}\delta_{lp}[\rho_{i}]_{qk}\overline{O_{i,lk}}O_{i,pq}.
\end{eqnarray*}
 Therefore we have derived 
\[
\sum_{i,j=1}^{M}\sum_{k,l=1}^{m_{i}}\sum_{p,q=1}^{m_{j}}\left[\delta_{ij}\delta_{lp}[\rho_{i}]_{qk}+(1-\delta_{ij})[\rho_{ij}]_{lq,kp}\right]\overline{O_{i,lk}}O_{j,pq}\geq0.
\]
 We can think of $O_{j,pq}$ as a vector $O\in\prod_{i=1}^{M}\mathbb{C}^{m_{i}\times m_{i}}\simeq\mathbb{C}^{\sum_{i=1}^{M}m_{i}^{2}}$.
The choice of such $O$ was completely arbitrary. Therefore we have
proved that the $\left(\sum_{i=1}^{M}m_{i}^{2}\right)\times\left(\sum_{i=1}^{M}m_{i}^{2}\right)$
matrix $G^{(2)}=G^{(2)}[\{\rho_{ij}\}_{i\leq j}]$ defined by 
\[
G_{ilk,jpq}^{(2)}:=\delta_{ij}\delta_{lp}[\rho_{i}]_{qk}+(1-\delta_{ij})[\rho_{ij}]_{lq,kp}
\]
 is positive semidefinite. This matrix can be thought of as a linear operator
$G^{(2)}:\prod_{i=1}^{M}\mathbb{C}^{m_{i}\times m_{i}}\ra\prod_{i=1}^{M}\mathbb{C}^{m_{i}\times m_{i}}$.
(One can readily check that $G^{(2)}$ is Hermitian.) For a quantum
spin system, we have $m_{i}=2$ for all $i$, so this is a semidefinite
constraint on a $(4M)\times(4M)$ matrix, which is can be efficiently
enforced.

At last we have derived a semidefinite relaxation, which we shall
call the \emph{two-marginal SDP}: 
\[
E_{0}^{(2)}=\inf_{\{\rho_{ij}\}_{i<j}\in\mathbf{LQM}_{2}(\mathcal{Q})\,:\,G^{(2)}[\{\rho_{ij}\}_{i\leq j}]\succeq0}\left(\sum_{i}\Tr\left[H_{i}\rho_{i}\right]+\sum_{i<j}\Tr\left[H_{ij}\rho_{ij}\right]\right).
\]
 The relaxation yields the energy lower bound $E_{0}\geq E_{0}^{(2)}$,
as well as a minimizer $\rho^{(2)}$ that is expected to approximate
the exact two-marginals.\\

The two-marginal SDP can be written, in expanded form, as 

\begin{eqnarray}
\underset{\{\rho_{i}\},\,\{\rho_{ij}\}_{i<j}}{\mbox{minimize}} \quad &  & \sum_{i}\Tr\left[H_{i}\rho_{i}\right]+\sum_{i<j}\Tr\left[H_{ij}\rho_{ij}\right]\label{eq:sdpObj}\\
\mbox{subject to} \quad &  & \rho_{ij}\succeq0,\quad i,j=1,\ldots,M,\label{eq:sdpC1}\\
 &  & \rho_{i}=\Tr_{\{j\}}[\rho_{ij}],\ \rho_{j}=\Tr_{\{i\}}[\rho_{ij}],\quad i,j=1,\ldots,M,\label{eq:sdpC2}\\
 &  & \Tr[\rho_{i}]=1,\quad i=1,\ldots,M,\label{eq:sdpC3}\\
 &  & G[\{\rho_{ij}\}_{i\leq j}]\succeq0.\label{eq:sdpC4}
\end{eqnarray}
 Although there are several ways to write constraints yielding the
same feasible set, the dual SDP is actually influenced by the choice
of constraints used to define this set. The choices made here will
yield interesting dual structure, to be explored in section~\ref{sec:duality}.

\subsection{Abstract perspective on the global semidefinite constraints}

\label{sec:abstractGlobal}More abstractly, it is useful to think
of $G=G[\{\rho_{ij}\}]$ as being composed of blocks $G_{ij}[\rho_{ij}]$
(indexed by marginal pairs $i,j$), defined by 
\[
\left(G_{ij}[\rho_{ij}]\right)_{\alpha\beta}=\begin{cases}
\Tr\left[\rho_{i}\,O_{i,\alpha}^{\dagger}O_{i,\beta}\right] & i=j\\
\Tr\left[\rho_{ij}\,O_{i,\alpha}^{\dagger}\otimes O_{j,\beta}\right] & i\neq j,
\end{cases}
\]
where $\left\{ O_{i,\alpha}\right\} _{\alpha=1}^{m_{i}^2}$ is basis for the
set of one-body operators on site $i$.  By considering $\alpha$ as a multi-index
$\alpha=(k,l)$ and choosing $\left(O_{i,(k,l)}\right)_{k',l'}=\delta_{kk'}\delta_{ll'}$
to be the `standard unit vectors' in $\mathbb{C}^{m_{i}\times m_{i}}$,
we exactly recover our former explicit representation of $G[\{\rho_{ij}\}]$.\\

\begin{rem}
(Restricted operator sets.)\label{rem:restrictedOpSets} The more
abstract perspective suggests a natural framework for further relaxation.
Suppose that for each $i=1,\ldots,M$, we are given a linearly independent
collection $\left\{ O_{i,\alpha}\right\} _{\alpha\in\mathcal{I}_{i}}$
of one-body operators for the $i$-th site, where $\mathcal{I}_{i}$
is a given index set. Then we can define $G=G[\rho^{(2)}]$ in terms
of blocks as above, where the block $G_{ij}[\rho_{ij}]$ is a matrix of size $\vert\mathcal{I}_{i}\vert\times\vert\mathcal{I}_{j}\vert$, defined once again by 
\[
\left(G_{ij}[\rho_{ij}]\right)_{\alpha\beta}=\begin{cases}
\Tr\left[\rho_{i}\,O_{i,\alpha}^{\dagger}O_{i,\beta}\right] & i=j\\
\Tr\left[\rho_{ij}\left(O_{i,\alpha}^{\dagger}\otimes O_{j,\beta}\right)\right] & i\neq j
\end{cases}
\]
 for $\alpha\in\mathcal{I}_{i}$, $\beta\in\mathcal{I}_{j}$. In principle
one can consider restricted index sets with $\vert\mathcal{I}_{i}\vert<m_{i}^{2}$
containing only the most physically important operators. Such restricted
structure will correspond to interesting structure from the perspective
of the dual problem to be considered below.
\end{rem}
$ $
\begin{rem}
\label{rem:quasilocal}
(Quasi-local constraints.) In order to improve the efficiency of the
semidefinite introduced above, one could enforce the semidefiniteness
of certain principal submatrices of $G$. For example, for each $k$, one
could define a submatrix $G^{(k)}$ of $G$ by restricting the block
indices $i,j$ to those satisfying $d(i,k),d(j,k)\leq d_{\max}$,
where $d(\cdot,\cdot)$ is an appropriate notion of distance between indices (e.g.,
graph distance for a lattice model) and $d_{\max}$ is a locality
parameter. Then one enforces $G^{(k)}[\{\rho_{ij}\}]\succeq0$ for
all $k$. For constant $d_{\max}$ suitably large, in principle such
constraints could achieve good performance while maintaining linear
scaling in $M$ of the SDP problem size for suitably local Hamiltonians,
by omitting $\rho_{ij}$ from the optimization variables for $d(i,j)>d_{\max}$.
\end{rem}

\subsection{Higher marginal constraints}
\label{sec:higherMarginal}
A tighter SDP relaxation can be derived by considering a set $\{\rho_{ijk}\}_{i<j<k}$
of quantum three-marginals as the optimization variable. One may enforce
the suitably defined local consistency constraints, denoted $\{\rho_{ijk}\}_{i<j<k}\in\mathbf{LQM}_{3}(\mathcal{Q})$,
then defining variables $\rho_{ij}$ in terms of the $\rho_{ijk}$
via partial traces, and additionally enforce $G[\{\rho_{ij}\}_{i\leq j}]\succeq0$.
We refer to the corresponding semidefinite relaxation as the \emph{three-marginal
SDP}.\\

To derive the corresponding semidefinite constraints, we have to keep track
of the four-marginals. Suitable necessary conditions can derived by
enforcing $\Tr\left[\rho\,\hat{O}^{\dagger}\hat{O}\right]\geq0$ for
all $\hat{O}$ of the form $\hat{O}=\sum_{i,i'}\hat{O}_{i,i'}$, where
the $\hat{O}_{i,i'}$ are \emph{two-body} operators. As such one may
define the \emph{four-marginal SDP}, and so on. Note that, e.g., the
four-marginal SDP can in fact accommodate more general Hamiltonians,
i.e., Hamiltonians including additional four-body terms.

\subsection{Cluster perspective}

\label{sec:cluster}In order to systematically improve the accuracy
of the two-marginal SDP, instead of considering higher marginals we
may alternately consider \emph{increasing the cluster size}. Formally,
such considerations will yield problems can still be accommodated
as special cases of our previously introduced setting. However, the
difference in perspective is noteworthy, and the generalization to
the case of overlapping clusters (considered in the next section)
is \emph{not }accommodated as such a special case.\\

Suppose that our site index set is written as a union of cluster index
sets $C_{\gamma}$, i.e., 
\[
\{1,\ldots,M\}=\bigcup_{\gamma=1}^{N_{\mathrm{c}}}C_{\gamma},
\]
 where the cluster index sets $C_{\gamma}$ are \emph{disjoint}. Then
one can define 
\[
Y_{\gamma}:=\prod_{i\in C_{\gamma}}X_{i}
\]
 to be the classical state space for the $\gamma$-th cluster. Then
by considering the clusters now as \emph{sites} with classical state
spaces $Y_{\gamma}$ and following the derivation of the two-marginal
SDP, we may derive the \emph{cluster two-marginal SDP}, relative to
the cluster decomposition $\{C_{\gamma}\}$. Note that this problem
may be viewed formally as a two-marginal SDP ; however, the distinction
makes sense when we think of the limit of expanding clusters for a
problem that is otherwise fixed. Higher-marginal cluster SDPs can
be derived similarly.

\subsection{Overlapping clusters}
\label{sec:overlapping}
We now demonstrate the treatment of overlapping clusters. Suppose again
that 
\[
\{1,\ldots,M\}=\bigcup_{\gamma=1}^{N_{\mathrm{c}}}C_{\gamma},
\]
 but now relax the assumption that the $C_{\gamma}$ are disjoint.
Since the overlap of two clusters might even be a single site of the
original model, we can no longer just `coarse-grain' clusters and
neglect all of their intra-cluster structure. In particular, imposition
of necessary local consistency constraints demands a bit more care.\\

Now the primary objects in our relaxation will be the two-\emph{cluster}
marginals, denoted $\rho_{\gamma\delta}:=\rho_{C_{\gamma}\cup C_{\delta}}$
for $\gamma\leq\delta$. Each $\rho_{\gamma\delta}$ is an operator
on the quantum state space specified by the \emph{union }of sites
$C_{\gamma}\cup C_{\delta}$, which may of course be smaller in size
than $\vert C_{\gamma}\vert+\vert C_{\delta}\vert$. Then the one-cluster
marginals $\rho_{\gamma}:=\rho_{C_{\gamma}}$ (which we sometimes
also denote by $\rho_{\gamma\gamma}$) are obtained in terms of the
two-cluster marginals via 
\[
\rho_{\gamma}=\Tr_{C_{\delta}\backslash C_{\gamma}}\left[\rho_{\gamma\delta}\right],\quad\rho_{\delta}=\Tr_{C_{\gamma}\backslash C_{\delta}}\left[\rho_{\gamma\delta}\right].
\]
 These identities yield consistency constraints analogous to the local
consistency constraints introduced earlier. However, we can also include
the \emph{overlap} constraints by introducing the variable $\rho_{(\gamma\delta)\cap(\gamma'\delta')}$
representing the marginal corresponding to the set $(C_{\gamma}\cup C_{\delta})\cap(C_{\gamma'}\cup C_{\delta'})$,
for all $\gamma<\delta$, $\gamma'<\delta'$, constrained by 
\[
\rho_{(\gamma\delta)\cap(\gamma'\delta')}=\Tr_{(C_{\gamma}\cup C_{\delta})\backslash(C_{\gamma'}\cup C_{\delta'})}\left[\rho_{\gamma\delta}\right]=\Tr_{(C_{\gamma'}\cup C_{\delta'})\backslash(C_{\gamma}\cap C_{\delta})}\left[\rho_{\gamma'\delta'}\right].
\]
 Note that these constraints are nontrivial only if the intersection
$(C_{\gamma}\cup C_{\delta})\cap(C_{\gamma'}\cup C_{\delta'})$ of
cluster pairs is nonempty.\\

To complete the discussion of the overlapping cluster two-marginal
SDP, we need to derive the semidefinite constraint. This is derived
by observing the necessary condition $\Tr\left[\rho\,\hat{O}^{\dagger}\hat{O}\right]\geq0$
for all $\hat{O}$ of the form $\hat{O}=\sum_{\gamma}\hat{O}_{\gamma}$
, where $\hat{O}_{\gamma}$ is a one-cluster operator, i.e., an operator
on $\bigotimes_{i\in C_{\gamma}}Q_{i}$, interpreted also (abusing
notation slightly) as an operator on $\mathcal{Q}$ by tensoring with
the identity on all sites outside of $C_{\gamma}$.\\

In fact, given a collection of one-cluster operators $\left\{ O_{\gamma,\alpha}\right\} _{\alpha\in\mathcal{I}_{\gamma}}$
for the $\gamma$-th cluster (i.e., operators on $\bigotimes_{i\in C_{\gamma}}Q_{i}$),
we build $G[\{\rho_{\gamma\delta}\}]$ blockwise by defining 
\[
\left(G_{\gamma\delta}[\rho_{\gamma\delta}]\right)_{\alpha\beta}=\Tr\left[\rho_{\gamma\delta}\,\tilde{O}_{\gamma,\alpha}^{\dagger}\tilde{O}_{\delta,\beta}\right]
\]
 for $\alpha\in\mathcal{I}_{\gamma}$, $\beta\in\mathcal{I}_{\delta}$,
and $\gamma<\delta$ (extending to $\gamma>\delta$ by hermiticity),
where $\tilde{O}_{\gamma,\alpha}$ is an operator on $\bigotimes_{i\in C_{\gamma}\cup C_{\delta}}Q_{i}$
obtained from $O_{\gamma,\alpha}$ by tensoring with the identity
operator over all sites in $C_{\delta}\backslash C_{\gamma}$. For
example, if $C_{\gamma}=\{1,2\}$ and $C_{\delta}=\{2,3\}$, then
we can represent $\tilde{O}_{\gamma,\alpha}=O_{\gamma,\alpha}\otimes I_{m_{3}}$
and $\tilde{O}_{\delta,\beta}=I_{m_{1}}\otimes O_{\delta,\beta}$
(recall that here $O_{\gamma,\alpha}$ is an operator on $Q_{1}\otimes Q_{2}$
and $O_{\delta,\beta}$ is an operator on $Q_{2}\otimes Q_{3}$).\\

The semidefinite constraint is, as before, $G[\{\rho_{\gamma\delta}\}]\succeq0$.
The resulting SDP can accommodate Hamiltonians of the form 
\[
\hat{H}=\sum_{\gamma}\hat{H}_{\gamma}+\sum_{\gamma<\delta}\hat{H}_{\gamma\delta},
\]
 where $\hat{H}_{\gamma}$ and $\hat{H}_{\gamma\delta}$ are one-cluster
and two-cluster operators, respectively.\\

Suitable analogous relaxations with higher overlapping cluster marginal
constraints may also be derived. We remark that the treatment of overlapping clusters here is significantly simpler and more principled than several other quantum embedding theories, including the dynamical mean-field theory (DMFT) and density matrix embedding theory (DMET).

\subsection{Translation-invariant setting}
\label{sec:translationInvariant}
In this section we describe how translation-invariant structure can
be exploited in a natural way in our semidefinite relaxation framework.
For simplicity we focus only on the case of the two-marginal SDP for
a translation-invariant Hamiltonian in one dimension. Extension to
higher dimensions is straightforward.\\

For the purposes of this section it is convenient to adopt a zero-indexing
convention for our site indices (usually denoted by $i,j$), i.e.,
we index our sites as $i=0,\ldots,M-1$. We obtain a translation-invariant
Hamiltonian by assuming that $\hat{H}_{i}=\hat{H}_{0}$ for all $i$
and $\hat{H}_{ij}=\hat{H}_{0,j-i}$ for all $i<j$. In turn we are
guaranteed translation-invariance of the ground-state density operator
(note: symmetry-breaking cannot occur for systems of finite size).
In particular, we have $\rho_{i}=\rho_{0}$ for all $i$ and $\rho_{ij}=\rho_{0,j-i}$
for all $i<j$, and it follows that we can constrain the matrix $G=G[\{\rho_{ij}\}]$
to be block-circulant, so that the block $G_{ij}$ depends only on
$i-j\ (\mathrm{mod}\ M)$. Hence all of the information of $G$ is
contained in the first row of blocks, and moreover $G$ can be block-diagonalized
by taking the blockwise discrete Fourier transform of the first row
of blocks. Indeed, these diagonal blocks are obtained as 
\[
\tilde{G}_{k}=\frac{1}{\sqrt{M}}\sum_{j=0}^{M-1}\exp\left(\imath\frac{2\pi jk}{M}\right)G_{0j},
\]
 $k=0,\ldots,M-1$, where we use `$\imath$' to denote the imaginary
unit to avoid confusion with our indexing notation. Now the constraint
$G\succeq0$ is equivalent to the constraint that $\tilde{G}_{k}\succeq0$
for all $k$. Hence we arrive at the periodic two-marginal SDP: 

\begin{eqnarray*}
\underset{\rho_{0},\,\{\rho_{0j}\}_{j=0,\ldots,M-1}}{\mbox{minimize}} \quad &  & \Tr\left[H_{0}\rho_{0}\right]+\sum_{j=1}^{M-1}\Tr\left[H_{0j}\rho_{0j}\right]\\
\mbox{subject to} \quad &  & \rho_{0j}\succeq0,\quad j=0,\ldots,M-1,\\
 &  & \rho_{0}=\Tr_{\{j\}}[\rho_{0j}],\ \rho_{0}=\Tr_{\{0\}}[\rho_{0j}],\quad j=0,\ldots,M-1,\\
 &  & \Tr[\rho_{0}]=1,\\
 &  & \sum_{j=0}^{M-1}\exp\left(\imath\frac{2\pi jk}{M}\right)G_{0j}[\rho_{0j}]\succeq0,\quad k=0,\ldots,M-1.
\end{eqnarray*}
 Notice that we have economized significantly on optimization variables,
and, moreover, we have exchanged a semidefinite constraint of size
$\sim M$ for $M$ semidefinite constraints of size constant in $M$.
Moreover, a careful implementation of a solver for this SDP should
be able to exploit the FFT in the implementation of the semidefinite
constraints.

\subsubsection{Periodicity constraints}

\label{sec:periodicityConstraints}If our sites are obtained as composite
sites representing non-overlapping clusters (as discussed in section
\ref{sec:cluster}) and if, moreover, our Hamiltonian is translation-invariant
with respect to these \emph{underlying} sites, then we can impose
further constraints to enforce the \emph{internal} translation-invariance
of our cluster marginals. To wit, in addition to our optimization
variables $\{\rho_{0\delta}^{\mathrm{C}}\}$ for the two-\emph{cluster}
marginals, we can define additional optimization variables $\{\rho_{0j}\}$
for the two-\emph{site} marginals and then enforce, for all $i\in C_{0},j\in\{1,\ldots,M\}$,
that $\rho_{0,j-i}=\Tr_{C_{0}\cup C_{\delta(j)}\backslash\{i,j\}}\left[\rho_{0,\delta(j)}^{\mathrm{C}}\right]$
, where $\delta(j)$ is the index of the cluster containing site $j$.
We refer to these additional constraints as \emph{periodicity constraints}.

\section{Fermions}

\label{sec:fermions}

\subsection{Preliminaries}
\label{sec:fermionPrelim}
The fundamental objects of fermionic systems in the second quantized formulation (see e.g.~\cite{NegeleOrland1988}) are the creation
operators $a_{1}^{\dagger},\ldots,a_{M}^{\dagger}$ and their Hermitian
adjoints, the annihilation operators $a_{i}$, which act on the Fock
space $\mathcal{F}$ and satisfy the canonical anticommutation relations 

\[
\{a_{i},a_{j}^{\dagger}\}=\delta_{ij},\quad\{a_{i},a_{j}\}=\{a_{i}^{\dagger},a_{j}^{\dagger}\}=0,
\]
 where $\{\,\cdot\,,\,\cdot\,\}$ denotes the anticommutator. One
defines the number operators by $\hat{n}_{i}:=a_{i}^{\dagger}a_{i}$
and the total number operator by $\hat{N}:=\sum_{i=1}^{M}\hat{n}_{i}$.\\

These objects can be concretely realized via the identification of
Hilbert spaces $\mathcal{F}\simeq\bigotimes^{M}\mathbb{C}^{2}\simeq\mathbb{C}^{2^{M}}$,
under which the annihilation operators correspond to quantum spin-$\tfrac{1}{2}$
operators as 
\[
a_{i}^{\dagger}\rightsquigarrow\underbrace{\sigma^{z}\otimes\cdots\otimes\sigma^{z}}_{i-1\,\mathrm{factors}}\otimes\left(\begin{array}{cc}
0 & 0\\
1 & 0
\end{array}\right)\otimes I_{2}\otimes\cdots\otimes I_{2}.
\]
 This identification of operators defines the Jordan-Wigner transformation
 (JWT)~\cite{NegeleOrland1988}. Note that the JWT depends on the ordering of the states in
the sense that permuting the states before the JWT is not equivalent
to permuting the tensor factors after the JWT. \\

After specifying a particle-number-conserving Hamiltonian $\hat{H}$,
i.e., a Hermitian operator on the Fock space which commutes with $\hat{N}$,
and a fixed particle number $N$, we are interested in computing the
$N$-particle ground state energy 
\[
E_{0}(N)=\inf\left\{ \langle\psi\vert\hat{H}\vert\psi\rangle\,:\,\vert\psi\rangle\in\mathcal{F},\,\langle\psi\vert\psi\rangle=1,\,\langle\psi\vert\hat{N}\vert\psi\rangle=N\right\} .
\]
 It is equivalent to solve 
\[
E_{0}(N)=\inf_{\rho\in\mathcal{D}(\mathcal{F})\,:\,\Tr[\hat{N}\rho]=N}\Tr\left[\hat{H}\rho\right],
\]
 where $\mathcal{D}(\mathcal{F})$ indicates the set of density operators
on the Fock space (i.e., positive semidefinite Hermitian operators
$\mathcal{F}\ra\mathcal{F}$ of unit trace).\\

Observe that although $\mathcal{F}$ can be identified with a quantum-spin
state space, the creation operators are \emph{not }one-qubit operators
in the sense of quantum spin systems, nor are hopping operators $a_{i}^{\dagger}a_{j}+a_{j}^{\dagger}a_{i}$
generically two-qubit operators. Moreover, the complexity of such
operators after the JWT can depend unphysically on the ordering of
the sites. Hence most second-quantized problems of interest (with
the exception of local one-dimensional models) \emph{simply do not
}fit into the framework of variational embedding introduced above
for quantum spin systems.\\

To illustrate this point and provide some concrete examples, we now
describe several Hamiltonians of interest in this setting. Of particular
note is the Hubbard model, whose states we enumerate via the orbital-spin
index $(i,\sigma)$, where $i=1,\ldots,M$, $\sigma=\uparrow,\downarrow$.
\begin{equation}
\hat{H}=-t\sum_{ij\sigma}A_{ij}a_{i\sigma}^{\dagger}a_{j\sigma}+U\sum_{i}\hat{n}_{i\uparrow}\hat{n}_{i\downarrow},\label{eq:hubbard}
\end{equation}
 where $A_{ij}$ is the adjacency matrix of a graph with vertex set
$\{1,\ldots,M\}$, e.g., a one-dimensional chain or a two-dimensional square lattice. The Hubbard model plays a significant
role in understanding strongly correlated quantum systems, such as the high
temperature superconductivity \cite{RaghuKivelsonScalapino2010}.
\\

More generally, one can consider a `generalized Coulomb model' of
the form 
\[
\hat{H}=\sum_{ij\sigma}h_{ij}a_{i\sigma}^{\dagger}a_{j\sigma}+\sum_{ij\sigma\tau}U_{ij}\hat{n}_{i\sigma}\hat{n}_{j\tau},
\]
 which includes in particular the Hubbard model and variants with
longer-range interactions. In fact, via certain choices of orbital
bases such as the planewave dual basis~\cite{MardirossianMcClainChan2018} and the recently introduced Gausslets~\cite{WhiteStoudenmire2019}, electronic
structure problems in the continuum can be mapped to second-quantized
Hamiltonians of this form. As we shall see, the generalized Coulomb
model is accommodated naturally within the framework of fermionic
variational embedding.\\

Broadening our view further still, consider a general two-body Hamiltonian
$\hat{H}$, written as 
\[
\hat{H}=\sum_{ij}h_{ij}a_{i}^{\dagger}a_{j}+\frac{1}{2}\sum_{ijkl}v_{ijkl}a_{i}^{\dagger}a_{j}^{\dagger}a_{l}a_{k}.
\]
Electronic structure problems in first quantization can be mapped to
such Hamiltonians via an arbitrary choice of orbital basis $\{\phi_{i}\}$
for (a subspace of) $L^{2}(\R^{d})$, where $d$ is the physical dimension.
If the basis functions have compact support, then $v_{ijkl}$ can
be nonzero only if both $\mathrm{\mathrm{supp}}(\phi_{i})\cap\mathrm{\mathrm{supp}}(\phi_{k})\neq\emptyset$
and $\mathrm{\mathrm{supp}}(\phi_{j})\cap\mathrm{\mathrm{supp}}(\phi_{l})\neq\emptyset$.
It will follow that after a suitable choice of overlapping clusters
(chosen so that each pair of intersecting basis functions ), such
Hamiltonians can also be accommodated within fermionic variational
embedding. We leave investigation of \emph{ab initio }quantum chemistry
problems by these means to future work.\\

In order to define a convex relaxation of the fermionic Gibbs variational
principle that is analogous to our relaxation for quantum spin systems,
we adopt a more abstract (and indeed general) perspective in section
\ref{sec:abstractFerm}, allowing for the derivation of a suitable
two-cluster-marginal SDP in \ref{sec:ferm2mar}. We will in fact see
in section \ref{sec:nonintExact} that our relaxation is tight for
noninteracting Hamiltonians, i.e., Hamiltonians that are quadratic
in the creation and annihilation operators. To our knowledge this feature has no analog
in the quantum spin setting because there is no related notion of
noninteracting systems. Then in section \ref{sec:concrete}, we will
describe how one can translate our abstract convex optimization problem
into an explicit SDP that can be implemented in practice.

\subsection{Abstract perspective}

\label{sec:abstractFerm}The fundamental objects of interest in the
abstract perspective is the \emph{algebra of operators }on the Fock
space. In fact, the Fock space itself plays no direct role in the
following developments, nor does any global JWT. Marginalization will
make use of the notion of a\emph{ subalgebra} subordinate to each
cluster. It is in the details of how these subalgebras lie within
the global algebra that the quantum-spin and fermionic cases differ.\\

Now let 
\[
\mathcal{A}:=\langle1,a_{1},\ldots,a_{M},a_{1}^{\dagger},\ldots,a_{M}^{\dagger}\rangle
\]
 denote the unital star-algebra over the complex numbers\footnote{A star-algebra over $\mathbb{C}$ is essentially an associative
algebra over $\mathbb{C}$ in which one can take adjoints, where the
adjoints satisfy their usual algebraic properties. `Unital' means that $1\in\mathcal{A}$. For further details, see, e.g.,~\cite{BratteliRobinson}. We will use no
deep results from the theory of star-algebras but nonetheless find
the perspective to be clarifying. In specific, it is useful to view
our algebra of fermionic operators independently from any Fock space
on which it acts, and in fact the notion of the Fock space does not
play any explicit role in our developments.} generated by the creation and annihilation operators subject to the
canonical anticommutation relations. (Throughout
we will use angle brackets to denote such generated algebras.) We
let $\hat{n}_{i}=a_{i}^{\dagger}a_{i}$ denote the corresponding number
operators and let $\hat{N}=\sum_{i}\hat{n}_{i}$ denote the total
number operator. Recall from above that for spinful models such as the Hubbard model, the
state index $i$ can be thought of as a composite orbital-spin index,
i.e., $i=(x,\sigma)$.\\

In fact the algebra $\mathcal{A}$ comes equipped with a $\mathbb{Z}_{2}$-grading,
i.e., we can write $\mathcal{A}$ as a direct sum of vector spaces
$\mathcal{A}=\mathcal{A}^{\mathrm{e}}\oplus\mathcal{A}^{\mathrm{o}}$,
where $\mathcal{A}^{\mathrm{e}}$ and $\mathcal{A}^{\mathrm{o}}$
denote the sets of even and odd operators, respectively. An operator
is even (resp., odd) if it can be written as a sum of even (resp.,
odd) monomials in $a_{1},\ldots,a_{M},a_{1}^{\dagger},\ldots,a_{M}^{\dagger}$.
(The reader can check that this notion is well-defined.) The $\mathbb{Z}_{2}$-grading
refers to the fact that $\mathcal{A}^{\mathrm{e}}\mathcal{A}^{\mathrm{e}}\subset\mathcal{A}^{\mathrm{e}}$,
$\mathcal{A}^{\mathrm{o}}\mathcal{A}^{\mathrm{o}}\subset\mathcal{A}^{\mathrm{e}}$,
$\mathcal{A}^{\mathrm{e}}\mathcal{A}^{\mathrm{o}}\subset\mathcal{A}^{\mathrm{o}}$,
and $\mathcal{A}^{\mathrm{o}}\mathcal{A}^{\mathrm{e}}\subset\mathcal{A}^{\mathrm{o}}$.\\

For any subset $C\subset\{1,\ldots,M\}$. Let $\mathcal{A}_{C}$ denote
the subalgebra 
\[
\mathcal{A}_{C}:=\left\langle \{1\}\cup\{a_{i},a_{i}^{\dagger}\,:\,i\in C\}\right\rangle ,
\]
 and let the even and odd components $\mathcal{A}_{C}^{\mathrm{e}}$
and $\mathcal{A}_{C}^{\mathrm{o}}$ be defined accordingly. Suppose
that our site index set is written as a union of cluster index sets
$C_{\gamma}$, i.e., 
\[
\{1,\ldots,M\}=\bigcup_{\gamma=1}^{N_{\mathrm{c}}}C_{\gamma},
\]
 where the cluster index sets $C_{\gamma}$ are disjoint, for simplicity.\\

We comment that, in contrast to our exposition for the case of quantum
spins, we shall directly work with general clusters (as opposed to
clusters consisting of a single site). The reason is that in the quantum
spin setting, it was possible to view non-overlapping clusters as
single sites (with enlarged local state spaces). Such a reduction
is not natural in the fermionic setting. Hence we retain the index
notation $\gamma,\delta$ for clusters and $i,j$ for individual sites
(of which the clusters are comprised).\\

We assume the Hamiltonian $\hat{H}\in\mathcal{A}$ can be written as a sum of one-cluster and two-cluster operators as 
\[
\hat{H}=\sum_{\gamma}\hat{H}_{\gamma}+\sum_{\gamma<\delta}\hat{H}_{\gamma\delta},
\]
 where $\hat{H}_{\gamma}\in\mathcal{A}_{C_{\gamma}}$ and $\hat{H}_{\gamma\delta}\in\mathcal{A}_{C_{\gamma}\cup C_{\delta}}$.\\

Note carefully for context that the subalgebra $\mathcal{A}_{C_{\gamma}}$
corresponds in our earlier setting of quantum spin systems to the
subalgebra of operators on $\bigotimes_{i\in C_{\gamma}}Q_{i}$, viewed
as operators on $\mathcal{Q}$ by tensoring with the identity. Clearly,
even by viewing the fermionic system as a spin system via JWT, this
subalgebra is inequivalent to the fermionic subalgebra above defined.
The reader should keep this perspective on the developments of section
\ref{sec:quantumSpins} in mind as we transpose them to the fermionic
setting.\\

Next we turn to defining our notion of a statistical ensemble and
its marginals. For this task we turn to the language of star-algebras.
The role of our full ensemble is played by the \emph{state}, a linear
functional $\omega:\mathcal{A}\ra\mathbb{C}$ such that $\omega(1)=1$
and $\omega(A^{\dagger}A)\geq0$ for any $A\in\mathcal{A}$. In our
setting (which is finite-dimensional), the action of a state can be
viewed as nothing more than tracing against a density operator on
the Fock space, as can be verified readily via the Riesz representation
theorem.
In the quantum spin setting of section \ref{sec:quantumSpins},
the action $\omega(\hat{A})$ of the state corresponds to the trace
$\Tr[\hat{A}\rho]$ against the density operator $\rho$. For $\hat{A}$
an operator on $\bigotimes_{i\in C}Q_{i}$, we have $\omega_{C}(\hat{A})=\omega(\hat{A})=\Tr[\hat{A}\rho]=\Tr[\hat{A}\rho_{C}]$,
i.e., our notion of marginalization---applied to a cluster subalgebra
in the quantum spin setting---precisely recovers the partial trace
operation. However, the abstract perspective will be useful in defining
the notion of a marginal because if we try to directly borrow the
corresponding notion from the setting of quantum spins, i.e., the
partial trace, then we find ourselves in need of a global JWT to proceed.\\

We let $\Omega$ denote the set of states on $\mathcal{A}$. 
Then in star-algebraic language, the $N$-particle ground-state energy
$E_{0}(N)$ minimization problem is naturally recast as 
\begin{equation}
E_{0}(N)=\inf_{\omega\in\Omega\,:\,\omega(\hat{N})=N}\omega(\hat{H}).\label{eq:fermGibbs}
\end{equation}

Next, our notion of a marginal in this setting is simply the restriction
of a state to a subalgebra. That is, for a subset $C\subset\{1,\ldots,M\}$,
we define the marginal $\omega_{C}$ via 
\[
\omega_{C}:=\omega\vert_{\mathcal{A}_{C}}.
\]
Of course, $\omega_{C}$ is itself a state on $\mathcal{A}_{C}$.
We let $\Omega_{C}$ denote the set of states on $\mathcal{A}_{C}$. Notice that,
as follows immediately from the definition, these sets are \emph{convex}.\\

\subsection{The two-cluster-marginal SDP}

\label{sec:ferm2mar}In this section we shall derive an `abstract
SDP' without describing how it can be realized on a computer. Later,
in section~\ref{sec:concrete}, we will describe how to achieve such realization (which
makes use of JWTs only for each pair of clusters). For simplicity,
we will only derive a relaxation that analogizes the \emph{(non-overlapping)
two-cluster-marginal SDP}. Further analogs can be derived by straightforward
(though perhaps tedious) modifications of the arguments presented
below.\\

For simplicity we denote the one-cluster marginals by $\omega_{\gamma}:=\omega_{C_{\gamma}}$
and the two-cluster marginals by $\omega_{\gamma\delta}:=\omega_{C_{\gamma}\cup C_{\delta}}$.
Note carefully from the definitions here that $\omega_{\gamma\delta}=\omega_{\delta\gamma}:\mathcal{A}_{C_{\gamma}\cup C_{\delta}}\ra\mathbb{C}$
and that $\omega_{\gamma\gamma}=\omega_{\gamma}:\mathcal{A}_{C_{\gamma}}\ra\mathbb{C}$.
Our one- and two-cluster marginals evidently satisfy the local consistency
constraints 
\[
\omega_{\gamma}=\omega_{\gamma\delta}\vert_{\mathcal{A}_{C_{\gamma}}},\quad\omega_{\delta}=\omega_{\gamma\delta}\vert_{\mathcal{A}_{C_{\delta}}}
\]
 via nested restriction operations.  By analogy to (\ref{eq:sum1bIneq}), our semidefinite constraint
will be derived from the observation that for any $\hat{A}\in\mathcal{A}$
of the form $\hat{A}=\sum_{\gamma}\hat{A}_{\gamma}$, where $\hat{A}_{\gamma}\in\mathcal{A}_{C_{\gamma}}$
for all $\gamma$, 
\[
0\leq\omega(\hat{A}^{\dagger}\hat{A})=\omega\left(\left[\sum_{\gamma}\hat{A}_{\gamma}\right]^{\dagger}\left[\sum_{\delta}\hat{A}_{\delta}\right]\right)=\sum_{\gamma\delta}\omega\left(\hat{A}_{\gamma}^{\dagger}\hat{A}_{\delta}\right).
\]
 Therefore the two-cluster marginals satisfy 
\[
\sum_{\gamma\delta}\omega_{\gamma\delta}\left(\hat{A}_{\gamma}^{\dagger}\hat{A}_{\delta}\right)\geq0
\]
 for all choices of $\{A_{\gamma}\}_{\gamma=1}^{N_{\mathrm{c}}}$
for which $\hat{A}_{\gamma}\in\mathcal{A}_{C_{\gamma}}$ for all $\gamma$.\\

More specifically, for each cluster $\gamma$ consider a \emph{list}
$\left\{ \hat{A}_{\gamma,\alpha}\right\} _{\alpha\in\mathcal{I}_{\gamma}}$
of operators in $\mathcal{A}_{C_{\gamma}}$, possibly (but not necessarily)
spanning the space of all operators in $\mathcal{A}_{C_{\gamma}}$.
(Compare to the perspective of section \ref{sec:abstractGlobal} on
the global semidefinite constraints in the quantum spin setting.)
Then one obtains $G\left[\{\omega_{\gamma\delta}\}\right]\succeq0$,
where $G=(G_{\gamma\delta})$ is specified blockwise by 
\[
\left(G_{\gamma\delta}[\omega_{\gamma\delta}]\right)_{\alpha\beta}=\omega_{\gamma\delta}\left(\hat{A}_{\gamma,\alpha}^{\dagger}\hat{A}_{\delta,\beta}\right).
\]
 In fact, $G=G[\{\omega_{\gamma\delta}\}_{\gamma\leq\delta}]$ depends
only on $\omega_{\gamma\delta}$ for $\gamma\leq\delta$ because the
lower triangular part can be obtained from the upper triangular part
via hermiticity.\\

Then we have derived the following relaxation of the variational principle
(\ref{eq:fermGibbs}), in which the $\omega_{\gamma}$ and $\omega_{\gamma\delta}$
are considered as optimization variables: 
\begin{eqnarray}
E_{0}^{(2)}(N)\ :=\ \ \underset{\{\omega_{\gamma}\},\,\{\omega_{\gamma\delta}\}_{\gamma<\delta}}{\mathrm{minimize}} \quad &  & \sum_{\gamma}\omega_{\gamma}\left(\hat{H}_{\gamma}\right)+\sum_{\gamma<\delta}\omega_{\gamma\delta}\left(\hat{H}_{\gamma\delta}\right),\label{eq:abstractSDP}\\
\mbox{subject to} \quad &  & \omega_{\gamma\delta}\in\Omega_{C_{\gamma}\cup C_{\delta}},\quad1\leq\gamma<\delta\leq N_{\mathrm{c}},\nonumber \\
 &  & \omega_{\gamma}=\omega_{\gamma\delta}\vert_{\mathcal{A}_{C_{\gamma}}},\quad\omega_{\delta}=\omega_{\gamma\delta}\vert_{\mathcal{A}_{C_{\delta}}},\quad1\leq\gamma<\delta\leq N_{\mathrm{c}},\nonumber \\
 &  & N=\sum_{\gamma}\omega_{\gamma}(\hat{N}_{\gamma}),\nonumber \\
 &  & G\left[\{\omega_{\gamma\delta}\}_{\gamma\leq\delta}\right]\succeq0,\nonumber 
\end{eqnarray}
 where $\hat{N}_{\gamma}:=\sum_{i\in C_{\gamma}}\hat{n}_{i}$ denotes
the $\gamma$-th cluster number operator. Since the constraints are
convex, we have specified an abstract convex optimization problem.
Now that we know that this relaxation makes sense in principle, our
hope is to express it later as a concrete semidefinite program.\\

It is computationally useful to realize a simplification. Physical
fermionic Hamiltonians are always even (including the anomalous, or
particle-number-\emph{non}conserving, Hamiltonians that arise in effective
descriptions of superconductivity), and hence one expects the action
of a physical state on an \emph{odd} operator in fact always yields
zero. Hence 
\[
\left(G_{\gamma\delta}[\omega_{\gamma\delta}]\right)_{\alpha\beta}=\omega_{\gamma\delta}\left(\hat{A}_{\gamma,\alpha}^{\dagger}\hat{A}_{\delta,\beta}\right)
\]
 is zero unless $\hat{A}_{\gamma,\alpha}$ and $\hat{A}_{\delta,\beta}$
are either both even or both odd. It follows that we can reduce the
size of the semidefinite constraint by splitting our operator lists
into even and odd subsets which we denote $\left\{ \hat{A}_{\gamma,\alpha}^{\mathrm{e}}\right\} _{\alpha\in\mathcal{I}_{\gamma}^{\mathrm{e}}}$
and $\left\{ \hat{A}_{\gamma,\alpha}^{\mathrm{o}}\right\} _{\alpha\in\mathcal{I}_{\gamma}^{\mathrm{o}}}$,
respectively. Then we define separate matrices $G^{\mathrm{e}}$ and
$G^{\mathrm{o}}$ blockwise by 
\begin{equation}
\left(G_{\gamma\delta}^{\mathrm{e/o}}[\omega_{\gamma\delta}]\right)_{\alpha\beta}=\omega_{\gamma\delta}\left(\left[\hat{A}_{\gamma,\alpha}^{\mathrm{e/o}}\right]^{\dagger}\left[\hat{A}_{\delta,\beta}^{\mathrm{e/o}}\right]\right).\label{eq:GeoBlock}
\end{equation}
 Then we may equivalently substitute our semidefinite constraint $G\succeq0$
with two semidefinite constraints $G^{\mathrm{e/o}}\succeq0$, each
of half (assuming that complete operator lists are chosen) the original
size.

\subsection{Exactness for noninteracting problems}\label{sec:exact_noninteracting}

\label{sec:nonintExact}In this section we assume that $\hat{H}$
is noninteracting, i.e., of the form $\hat{H}=\sum_{ij}{h_{ij}}\,a_{i}^{\dagger}a_{j}$,
where $h=(h_{ij})$ is Hermitian. We want to show that in this setting
$E_{0}^{(2)}(N)=E_{0}(N)$, i.e., the relaxation just introduced is
tight, under the meager further assumption that for each $i\in\{1,\ldots,M\}$,
the operators $a_{i},a_{i}^{\dagger}$ are contained in some cluster's
operator list.\\

Indeed, under this latter assumption it is not hard to see that the
matrices $D(\omega_{\{i,j\}}):=\left(\omega_{\{i,j\}}(a_{i}^{\dagger}a_{j})\right)_{i,j=1}^{M}$
and $D'(\omega_{\{i,j\}}):=\left(\omega_{\{i,j\}}(a_{i}a_{j}^{\dagger})\right)_{i,j=1}^{M}$
appear as principal submatrices of $G^{\mathrm{o}}[\{\omega_{\gamma\delta}\}]$,
where the two-site marginals $\omega_{\{i,j\}}$ are suitably obtained
in terms of the two-cluster marginals $\omega_{\gamma\delta}$ by appropriate
restriction. Note that by the fermionic anticommutation relations,
in fact $D'(\omega_{\{i,j\}})=I_{M}-D(\omega_{\{i,j\}})^{\top}$. Hence for
any feasible solution to our SDP, we have $0\preceq D(\omega_{\{i,j\}})\preceq I_{M}$.
Then it follows that $E_{0}^{(2)}(N)$ is an upper bound for
the optimal value $E'_{0}(N)$ of the following (further relaxed)
SDP: 
\begin{eqnarray*}
E'_{0}(N)\ :=\ \ \underset{D\in\mathbb{C}^{M\times M}}{\mathrm{minimize}} \quad &  & \Tr[D^{\top} h]\\
\mbox{subject to} \quad &  & 0\preceq D\preceq I_{M},\\
 &  & \Tr[D]=N.
\end{eqnarray*}
On the other hand, $E'_{0}(N)=\sum_{i=1}^{N}\lambda_{i}(h),$
where $\lambda_{i}(h)$ indicates the $i$-th lowest eigenvalue of
$h$. For noninteracting problems this is precisely the value of $E_{0}(N)$.
Hence we have shown $E_{0}(N)\geq E_{0}^{(2)}(N)\geq E'_{0}(N)=E_{0}(N)$,
from which it follows that $E_{0}^{(2)}(N)=E_{0}(N)$.\\

For certain problems, one may also expect asymptotic tightness in the limit
of strong interaction. For example, in the $t\ra0$ (or equivalently, $U\ra \infty$) limit of the Hubbard
model, the sites completely decouple, and it can be checked readily
that our SDP is tight in this scenario.

\subsection{Concrete perspective} \label{sec:concrete}

In order to represent $\omega_{\gamma\delta}$ in concrete terms, note that $\omega_{\gamma\delta}$ is defined by
its action on $\mathcal{A}_{C_{\gamma}\cup C_{\delta}}$. It is at
this point that we introduce for computational purposes the JWT, though
only for restricted fermionic algebras. Let $\mathrm{End}(V)$ denote the set of all endomorphisms of $V$. After specifying ordering
the sites of $C_{\gamma}\cup C_{\delta}$, i.e., a labeling map $\kappa_{\gamma\delta}:C_{\gamma}\cup C_{\delta}\ra\{1,\ldots,L_{\gamma\delta}\}$
where $L_{\gamma\delta}:=\vert C_{\gamma}\cup C_{\delta}\vert$, the
corresponding JWT fixes an algebra isomorphism $\mathcal{J}_{\gamma\delta}:\mathcal{A}_{C_{\gamma}\cup C_{\delta}}\ra\mathrm{End}\left(\bigotimes_{i=1}^{L_{\gamma\delta}}\mathbb{C}^{2}\right)$,
and we define $c_{\kappa(i)}^{\gamma\delta}\in\mathrm{End}\left(\bigotimes_{i=1}^{L_{\gamma\delta}}\mathbb{C}^{2}\right)$
to be the image of $a_{i}$ under this isomorphism for $i\in C_{\gamma}\cup C_{\delta}$.
More specifically, the transformation $\mathcal{J}_{\gamma\delta}$
is specified by setting $\mathcal{J}_{\gamma\delta}(a_{\kappa_{\gamma\delta}^{-1}(i)})=c_{i}^{\gamma\delta}$,
where 
\[
c_{i}^{\gamma\delta}:=\underbrace{\sigma^{z}\otimes\cdots\otimes\sigma^{z}}_{(i-1)\,\mathrm{factors}}\otimes\left(\begin{array}{cc}
0 & 1\\
0 & 0
\end{array}\right)\otimes\underbrace{I_{2}\otimes\cdots\otimes I_{2}}_{(L_{\gamma\delta}-i)\,\mathrm{factors}}.
\]
Notice that the case $\gamma=\delta$ makes perfect sense according
to the above definitions, though we will also introduce the alternative
notation $\mathcal{J}_{\gamma}:=\mathcal{J}_{\gamma\gamma}$.\\

Let $\mathrm{Id}\in \mathrm{End}\left(\bigotimes_{i=1}^{L_{\gamma\delta}}\mathbb{C}^{2}\right)$ be the identity operator. Then
$F_{\gamma\delta}:=\mathcal{J}_{\gamma\delta}\circ\omega_{\gamma\delta}\circ\mathcal{J}_{\gamma\delta}^{-1}$
is a linear functional on $\mathrm{End}\left(\bigotimes_{i=1}^{L_{\gamma\delta}}\mathbb{C}^{2}\right)$
satisfying $F_{\gamma\delta}(\mathrm{Id})=1$  and $F_{\gamma\delta}(A^{\dagger}A)\geq0$
for any $A\in\mathrm{End}\left(\bigotimes_{i=1}^{L_{\gamma\delta}}\mathbb{C}^{2}\right)$.
It follows (via the Riesz representation theorem) that there exists a unique $\rho_{\gamma\delta}\succeq0$
with $\Tr[\rho_{\gamma\delta}]=1$ such that $F_{\gamma\delta}(A)=\Tr[A\rho_{\gamma\delta}]$
for all $A\in\mathrm{End}\left(\bigotimes_{i=1}^{L_{\gamma\delta}}\mathbb{C}^{2}\right)$.
That is to say, $\omega_{\gamma\delta}(\hat{A})=\Tr[A\rho_{\gamma\delta}]$
whenever $A=\mathcal{J}_{\gamma\delta}(\hat{A})$. Again, we introduce
the alternative notation $\rho_{\gamma}=\rho_{\gamma\gamma}$ for
conceptual clarity.\\

Motivated by the preceding, we shall replace optimization over \emph{states}
$\omega_{\gamma\delta}:\mathcal{A}_{C_{\gamma}\cup C_{\delta}}\ra\mathbb{C}$
with optimization over \emph{density operators} $\rho_{\gamma\delta}\in\mathrm{End}\left(\bigotimes_{i=1}^{L_{\gamma\delta}}\mathbb{C}^{2}\right)$.
Crucially, the correspondence between states and density operators
has relied on a separate JWT for \emph{each }pair $(\gamma,\delta)$,
not a single global JWT that maps the global fermionic state to a
global density operator. Neither should we obtain $\rho_{\gamma\delta}$ from a global density operator $\rho$ via the standard definition of the partial trace, as in the case of quantum spin systems.

Under this correspondence $G_{\gamma\delta}^{\mathrm{e/o}}[\omega_{\gamma\delta}]$
as defined (\ref{eq:GeoBlock}) can be obtained as 

\[
\left(G_{\gamma\delta}^{\mathrm{e/o}}[\rho_{\gamma\delta}]\right)_{\alpha\beta}=\Tr\left(\left[\mathcal{J}_{\gamma\delta}\left(\hat{A}_{\gamma,\alpha}^{\mathrm{e/o}}\right)\right]^{\dagger}\left[\mathcal{J}_{\gamma\delta}\left(\hat{A}_{\delta,\beta}^{\mathrm{e/o}}\right)\right]\rho_{\gamma\delta}\right),
\]
 where we abuse notation slightly by identifying $G_{\gamma\delta}^{\mathrm{e/o}}[\rho_{\gamma\delta}]$
with $G_{\gamma\delta}^{\mathrm{e/o}}[\omega_{\gamma\delta}]$.\\

In order to write down a concrete realization of the optimization
problem (\ref{eq:abstractSDP}), the hurdle that remains is to encode
the local consistency constraints $\omega_{\gamma}=\omega_{\gamma\delta}\vert_{\mathcal{A}_{C_{\gamma}}}$
and $\omega_{\delta}=\omega_{\gamma\delta}\vert_{\mathcal{A}_{C_{\delta}}}$
for $\gamma<\delta$, which require us to further `marginalize' our fermionic
states.\\

To see how to do this, we first assume that the labeling map $\kappa_{\gamma\delta}$
satisfies $\kappa_{\gamma\delta}(C_{\gamma})<\kappa_{\gamma\delta}(C_{\delta})$
in the sense that every element of the left-hand side is less than
every element of the right-hand side. In the case of overlapping
clusters, which (as previously mentioned) we shall not discuss in full detail, the relevant generalization ensures that $\kappa_{\gamma\delta}(C_{\gamma})<\kappa_{\gamma\delta}([C_{\gamma}\cup C_{\delta}]\backslash C_{\gamma})$. For simplicity we also assume that $\kappa_{\gamma\delta}\vert_{C_{\gamma}}=\kappa_{\gamma\gamma}$
for all $\gamma<\delta$, and from now on we think of the labeling
maps $\kappa_{\gamma\delta}$ as fixed. It is always possible to choose
a labeling that satisfies these assumptions.\\

Then it follows from the definition of the JWT that for any $\hat{A} \in \mathcal{A}_{C_\gamma}$, $A:=\mathcal{J}_{\gamma\delta}(\hat{A})$ is of
the form 
\[
A=B\otimes\mathrm{Id}_{\bigotimes_{i=1}^{\vert C_{\delta}\vert}\mathbb{C}^{2}}=B\otimes\underbrace{I_{2}\otimes\cdots\otimes I_{2}}_{\vert C_{\delta}\vert\,\mathrm{factors}},
\]
 where $B=\mathcal{J}_{\gamma}(\hat{A})\in\mathrm{End}\left(\bigotimes_{i=1}^{L_{\gamma}}\mathbb{C}^{2}\right)$.
Then 
\[
\omega_{\gamma\delta}(\hat{A})=\Tr[A\rho_{\gamma\delta}]=\Tr[B\tilde{\rho}_{\gamma}],
\]
 where $\tilde{\rho}_{\gamma}:=\Tr_{\kappa_{\gamma\delta}(C_{\gamma})}[\rho_{\gamma\delta}]$.
Meanwhile, we have $\mathcal{J}_{\gamma}(\hat{A})=B$, and $\omega_{\gamma}(\hat{A})=\Tr[B\rho_{\gamma}]$.
Hence the constraint $\omega_{\gamma}=\omega_{\gamma\delta}\vert_{\mathcal{A}_{C_{\gamma}}}$
for $\gamma<\delta$ is equivalent to the stipulation that $\Tr[B\tilde{\rho}_{\gamma}]=\Tr[B\rho_{\gamma}]$
for all $B$, i.e., that 
\[
\rho_{\gamma}=\Tr_{\kappa_{\gamma\delta}(C_{\delta})}[\rho_{\gamma\delta}].
\]
Here $\Tr_{\kappa_{\gamma\delta}(C_{\delta})}(\cdot)$ \textit{is} the standard partial trace.

Meanwhile, for any $A=\mathcal{J}_{\gamma\delta}(\hat{A})$ where
$\hat{A}\in\mathcal{A}_{C_{\delta}}^{\mathrm{e}}$ is \emph{even},
we can write 
\[
A=\underbrace{I_{2}\otimes\cdots\otimes I_{2}}_{\vert C_{\gamma}\vert\,\mathrm{factors}}\otimes B,
\]
 where $B=\mathcal{J}_{\delta}(\hat{A})\in\mathrm{End}\left(\bigotimes_{i=1}^{L_{\delta}}\mathbb{C}^{2}\right)$.
Hence for all $\hat{A}\in\mathcal{A}_{C_{\delta}}^{\mathrm{e}}$,
we derive as above that $\omega_{\delta}(\hat{A})=\Tr\left[B\tilde{\rho}_{\delta}\right]$,
where $\tilde{\rho}_{\delta}:=\Tr_{\kappa_{\gamma\delta}(C_{\delta})}[\rho_{\gamma\delta}]$.
But for $\hat{A}\in\mathcal{A}_{C_{\delta}}^{\mathrm{o}}$, as mentioned
above we can assume $\omega_{\gamma\delta}(\hat{A})=\omega_{\delta}(\hat{A})=0$
(because this identity is a necessary condition satisfied by the exact
marginals) and hence also that $\Tr[B\rho_{\delta}]=0=\Tr[B\tilde{\rho}_{\delta}]$
for all $B\in\mathcal{J}_{\delta}(\mathcal{A}_{C_{\delta}}^{\mathrm{o}})$.
Thus the constraint $\omega_{\delta}=\omega_{\gamma\delta}\vert_{\mathcal{A}_{C_{\delta}}}$
for $\gamma<\delta$ is equivalent to the stipulation that $\Tr\left[B\tilde{\rho}_{\delta}\right]=\Tr[B\rho_{\delta}]$
for all $B$, i.e., that 
\[
\rho_{\delta}=\Tr_{\kappa_{\gamma\delta}(C_{\delta})}[\rho_{\gamma\delta}].
\]
Finally, note that the constraint $\Tr[\rho_{\gamma\delta}]=1$ can
simply be encoded, given our first local consistency constraint, by
$\Tr[\rho_{\gamma}]=1$. Then we obtain the following concrete realization
of (\ref{eq:abstractSDP}): 

\begin{eqnarray*}
E_{0}^{(2)}(N)\ :=\ \ \underset{\{\rho_{\gamma}\},\,\{\rho_{\gamma\delta}\}_{\gamma<\delta}}{\mathrm{minimize}} \quad &  & \sum_{\gamma}\Tr\left[\mathcal{J}_{\gamma}\left(\hat{H}_{\gamma}\right)\rho_{\gamma}\right]+\sum_{\gamma<\delta}\Tr\left[\mathcal{J}_{\gamma\delta}\left(\hat{H}_{\gamma\delta}\right)\rho_{\gamma\delta}\right],\\
\mbox{subject to} \quad &  & \rho_{\gamma\delta}\succeq0,\quad1\leq\gamma<\delta\leq N_{\mathrm{c}},\\
 &  & \rho_{\gamma}=\Tr_{\kappa_{\gamma\delta}(C_{\delta})}[\rho_{\gamma\delta}],\quad\rho_{\delta}=\Tr_{\kappa_{\gamma\delta}(C_{\gamma})}[\rho_{\gamma\delta}],\quad1\leq\gamma<\delta\leq N_{\mathrm{c}},\\
 &  & \Tr[\rho_{\gamma}]=1,\quad\gamma=1,\ldots,N_{\mathrm{c}},\\
 &  & N=\sum_{\gamma}\Tr\left[\mathcal{J}_{\gamma}\left(\hat{N}_{\gamma}\right)\rho_{\gamma}\right],\\
 &  & G\left[\{\rho_{\gamma\delta}\}_{\gamma\leq\delta}\right]\succeq0.
\end{eqnarray*}

\section{Numerical results}
\label{sec:numerics}
All numerical results were computed in $\mathsf{MATLAB}$ with $\mathsf{CVX}$~\cite{grant2008cvx} 
for performing SDP calculations. We limit our experiments to problems that are
small enough to validate by exact diagonalization. In particular,
we will illustrate numerically the fact that all of our relaxations
must yield lower bounds for the exact energy. We will also show that the omission of the global semidefinite 
constraints results in looser lower bounds, i.e., the global semidefinite constraints are nontrivial, 
even though the Hamiltonians are all local.
 As discussed in section~\ref{sec:compPartialDual} below, 
a more scalable implementation should be possible,
but such an implementation (as well as an accompanying numerical study
of properties of larger systems, e.g., approaching a thermodynamic
limit) will be left to future work.

\subsection{Transverse-field Ising model}
\label{sec:tfi}
\begin{figure}
\centering{}\includegraphics[bb=40bp 0bp 520bp 420bp,scale=0.40]{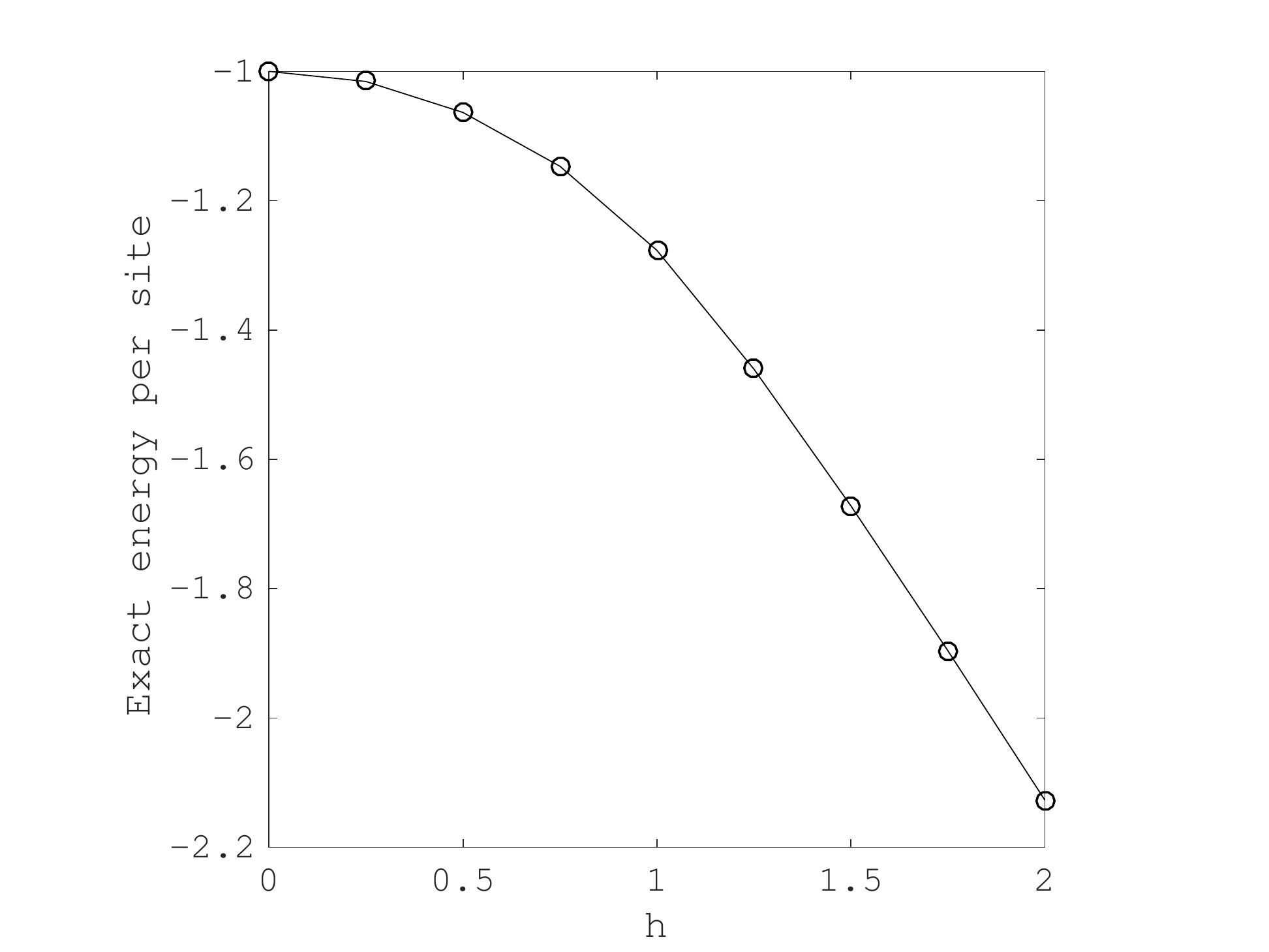}\includegraphics[bb=40bp 0bp 520bp 420bp,scale=0.40]{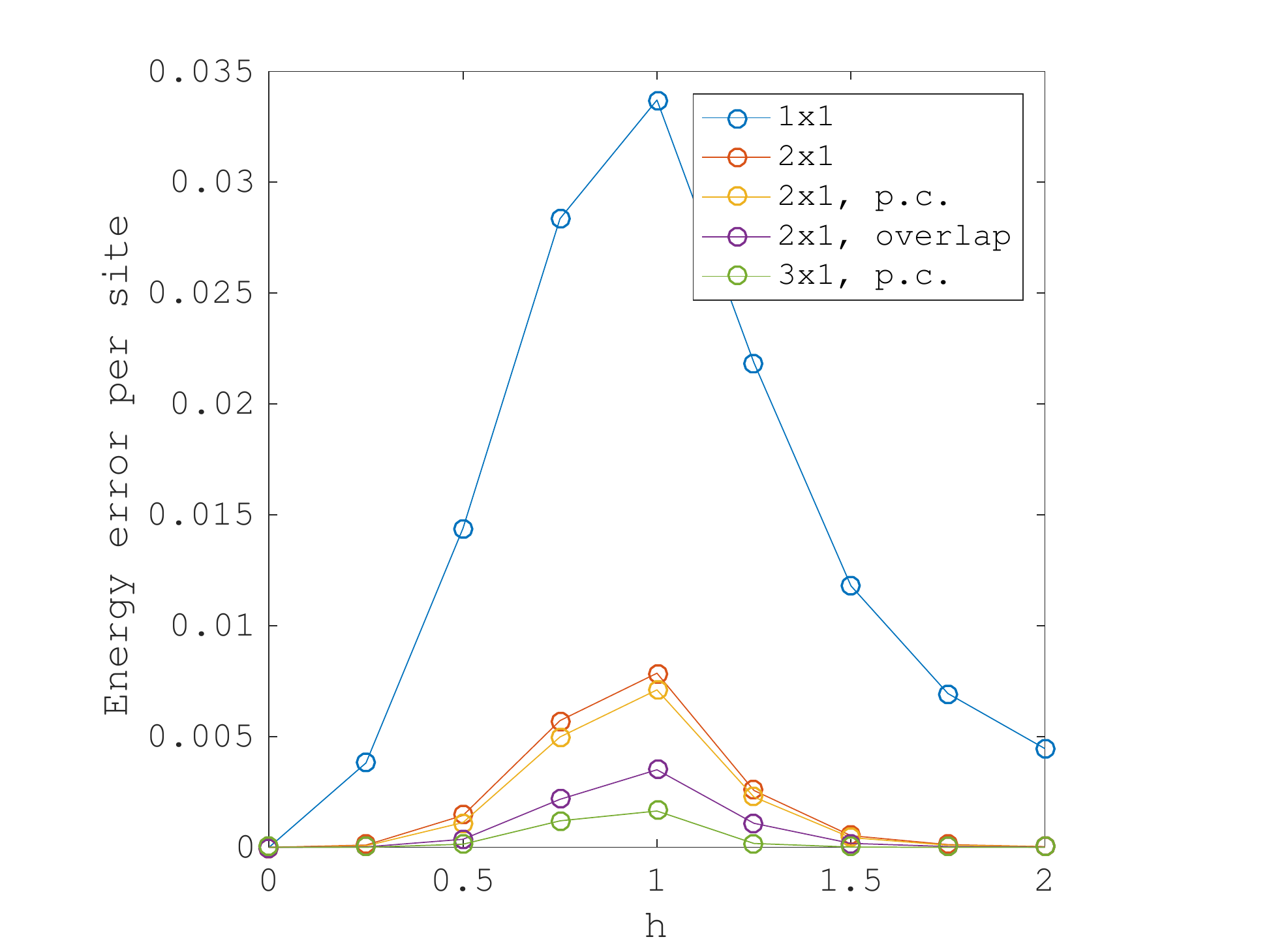}\caption{TFI model on periodic $12\times1$ lattice. Approximate energies are
computed via the two-cluster-marginal relaxation. Note that `p.c.'
indicates the inclusion of the periodicity constraints introduced
in section \ref{sec:periodicityConstraints}, and `overlap' indicates
that the choice of overlapping $2\times1$ clusters, i.e., $\{1,2\},\{2,3\},\{3,4\},\ldots,\{11,12\},\{12, 1\}$.\label{fig:tfi_12x1}}
\end{figure}

First we consider the transverse-field Ising (TFI) model (\ref{eq:tfi})
on a periodic $12\times1$ lattice, comparing results of the
two-cluster-marginal SDP for various cluster sizes. We also test the
periodicity constraints of section~\ref{sec:periodicityConstraints} and the
case of overlapping clusters. The results are shown in Figure \ref{fig:tfi_12x1}.

Note that, as the theory requires, all approximations do indeed yield lower bounds 
for the exact energy. Moreover these bounds become tighter for larger cluster sizes. 
Also notice that the case of overlapping $2\times 1$ clusters compares favorably to the 
case of non-overlapping $2 \times 1$ clusters, achieving an energy error roughly twice as small. 
(In the case of overlapping clusters, the periodicity constraints of section~\ref{sec:periodicityConstraints} 
are satisfied automatically by the solution, and there is no need to enforce them explicitly. Hence from 
Figure \ref{fig:tfi_12x1} it is clear that most of the improvement yielded by allowing for overlap  
is \emph{not} merely due to these constraints.)

\begin{figure}[ht]
\centering
\includegraphics[bb=40bp 0bp 520bp 420bp,scale=0.40]{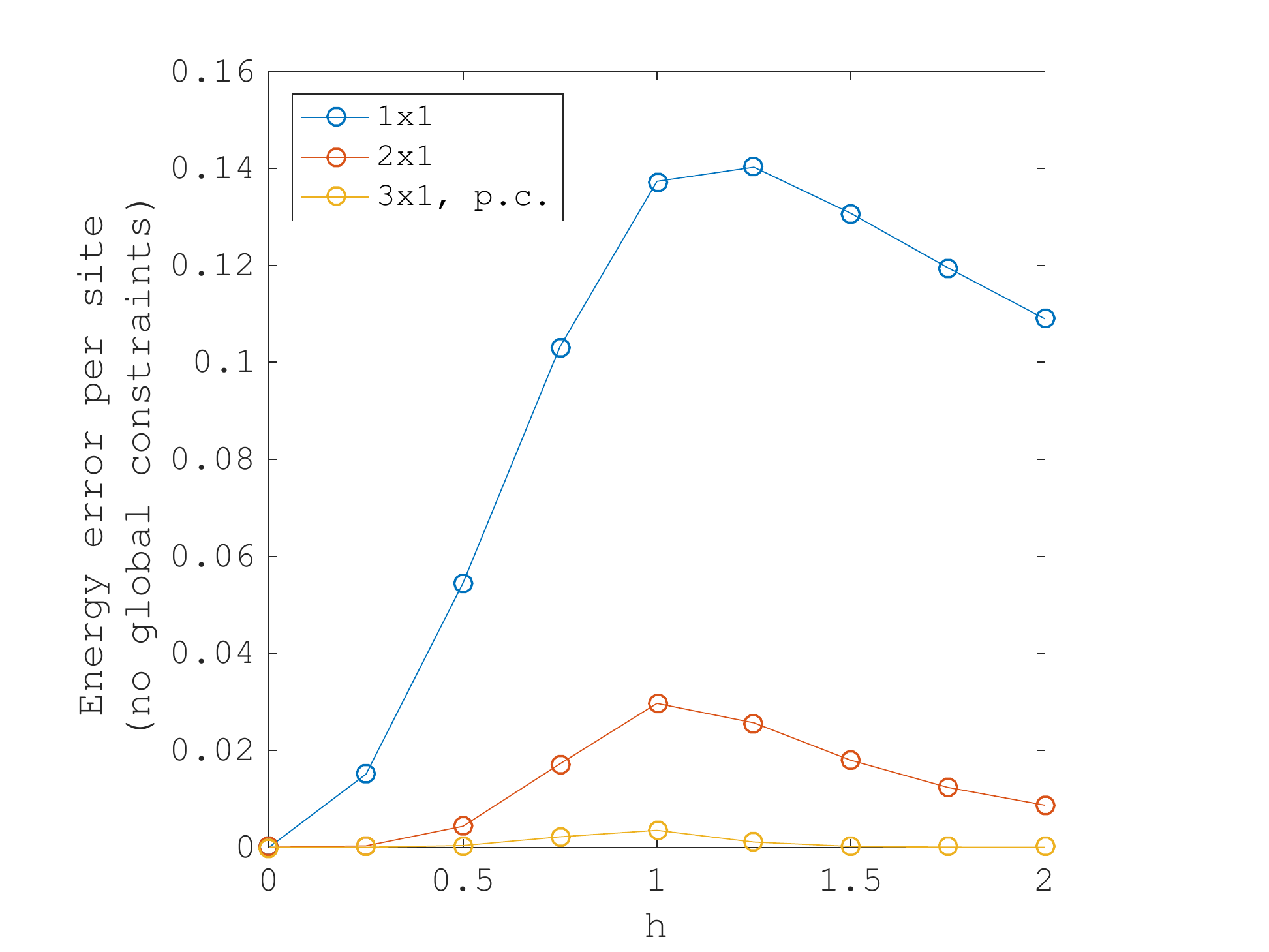}
\caption{Results for the same model and same relaxations as in Figure \ref{fig:tfi_12x1}, with the modification that the 
global semidefinite constraints are omitted in all cases. In this experiment the curves for `$2\times 1$, p.c.' and $2\times 1$, overlap' coincide with that of `$2\times 1$.' Note the change of scale of the vertical axis relative to the analogous plot of Figure \ref{fig:tfi_12x1}. For clarity, we remark that the value of the `$3\times1$, p.c.' curve at $h=1$ is 0.0035, compared to the corresponding value (with global constraints active) of 0.0016 depicted in Figure \ref{fig:tfi_12x1}.}
\label{fig:tfi_12x1_noGC}
\end{figure}

In Figure \ref{fig:tfi_12x1_noGC} we test the same relaxations on the same model problem, except that 
we \emph{omit the global semidefinite constraints}.  Neglecting the global semidefinite constraints correspond to the use of belief propagation (BP)~\cite{Pearl1982} in the classical setting, and its quantum generalization~\cite{LeiferPoulin2008,PoulinHastings2011,BarthelHubener2012,FerrisPoulin2013}.
Note that the omission of these constraints results 
in a significant degradation of the lower bound, even though the Hamiltonian is local.

\begin{figure}[ht]
\centering{}\includegraphics[bb=40bp 0bp 520bp 420bp,scale=0.40]{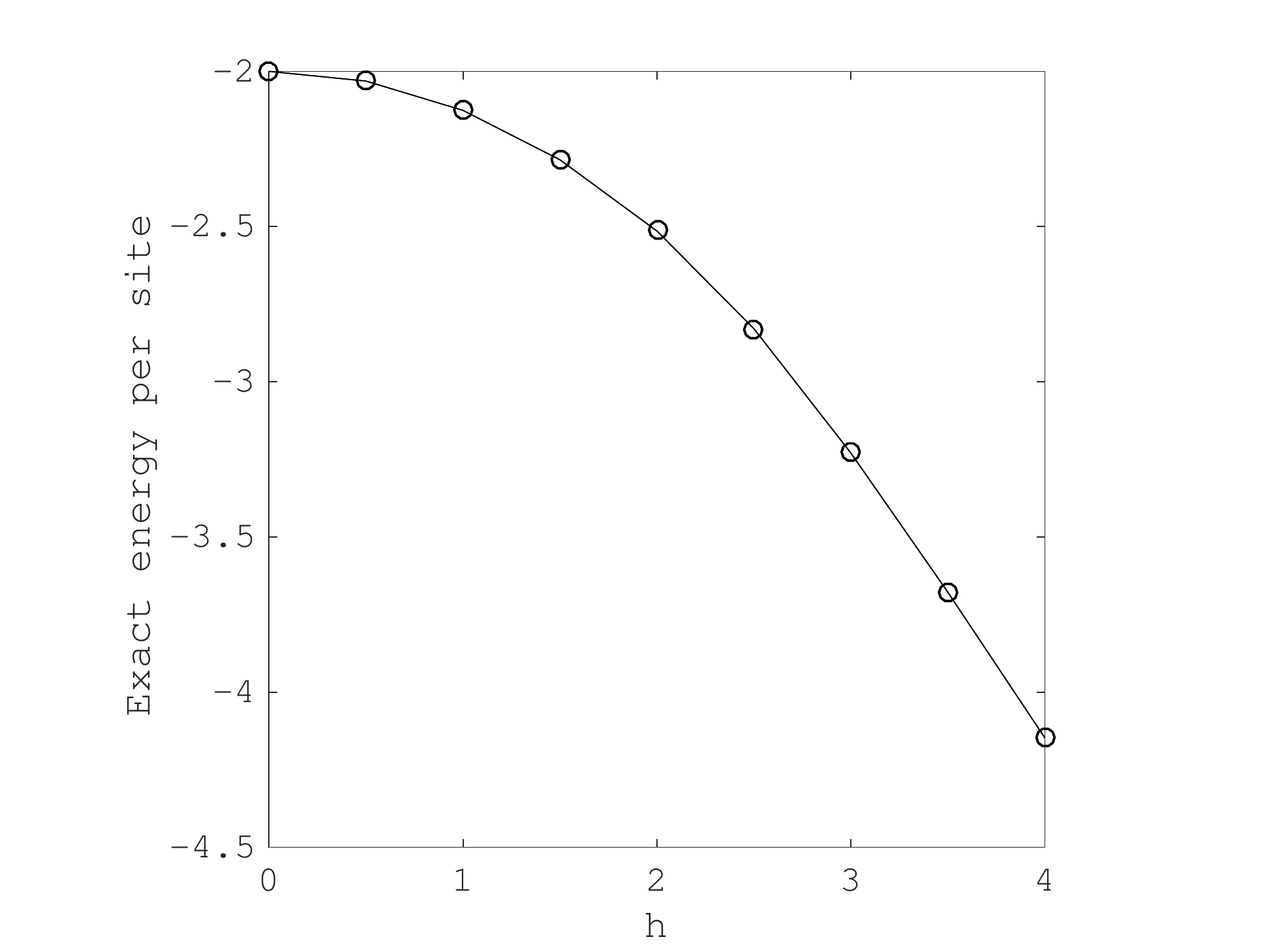}
\includegraphics[bb=40bp 0bp 520bp 420bp,scale=0.40]{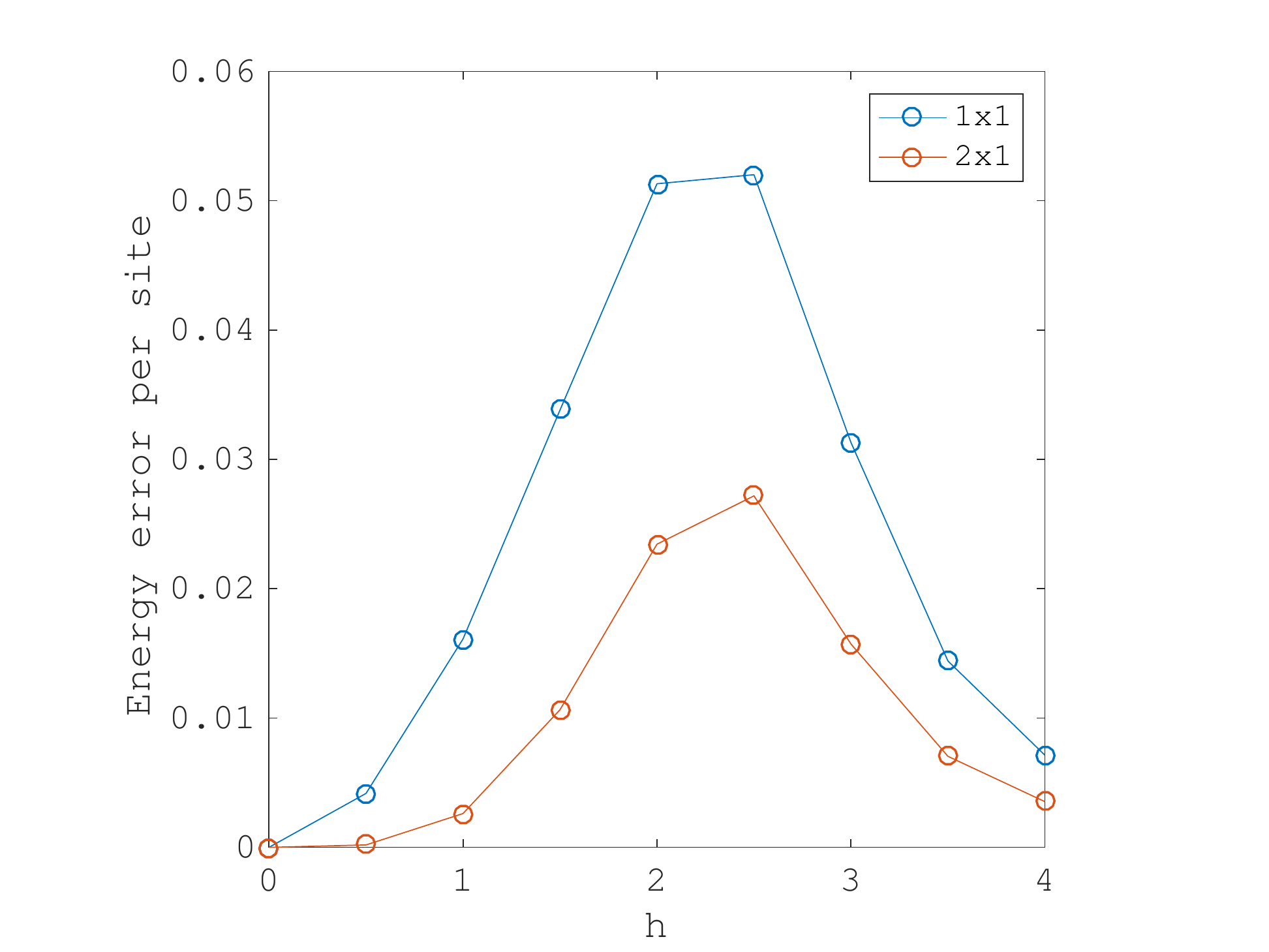}
\caption{TFI model on periodic $4\times3$ lattice. Approximate energies are
computed via the two-cluster-marginal relaxation.}
\label{fig:tfi_4x3}
\end{figure}

\begin{figure}[ht]
\centering{}\includegraphics[bb=40bp 0bp 520bp 420bp,scale=0.40]{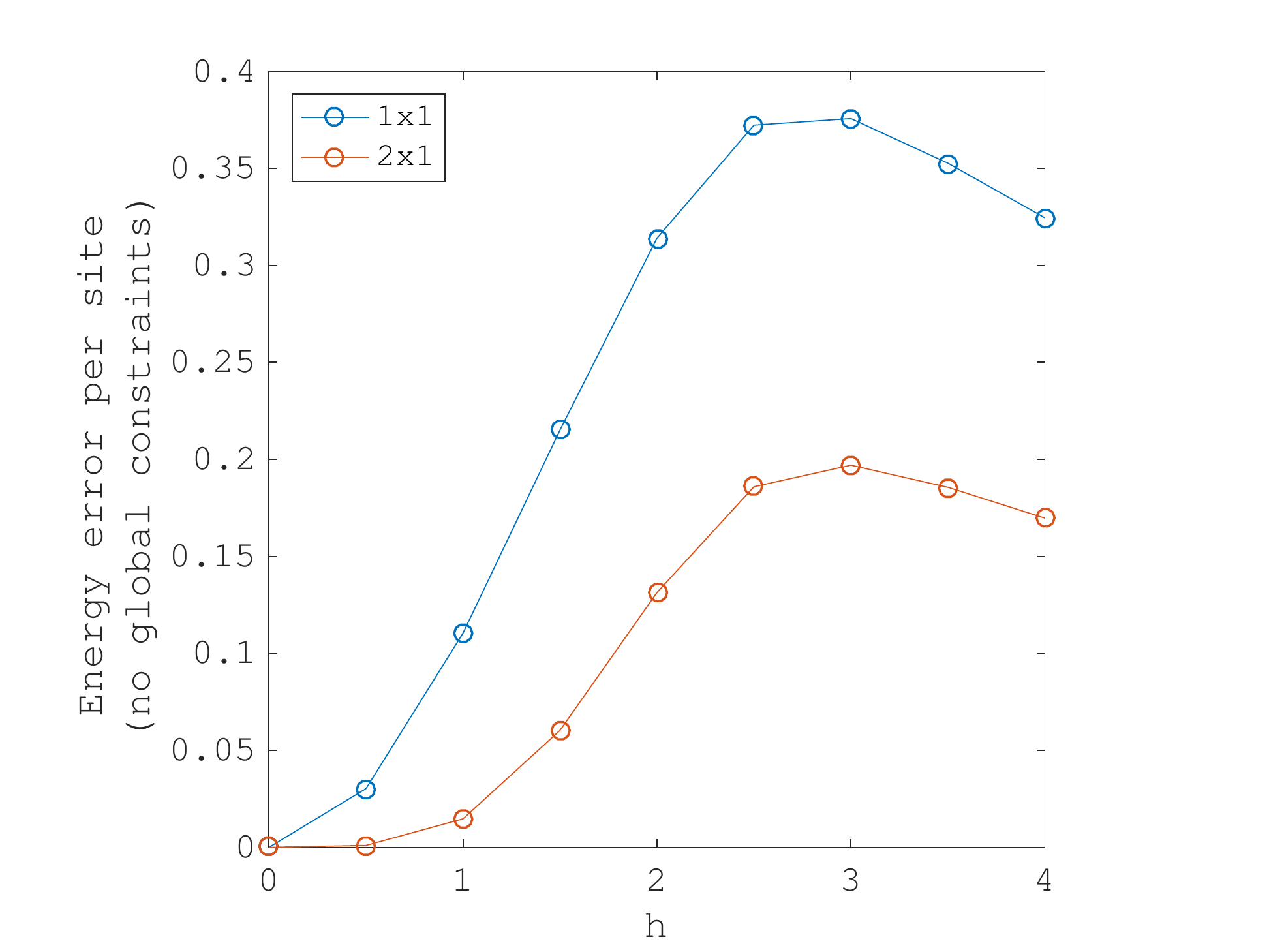}\caption{Results for the same model and same relaxations as in Figure \ref{fig:tfi_4x3}, with the modification that the 
global semidefinite constraints are omitted in all cases. Note the change of scale of the vertical axis relative to the analogous plot of Figure \ref{fig:tfi_4x3}.}\label{fig:tfi_4x3_noGC}
\end{figure}

 Next we consider the TFI model on a periodic $4\times3$ square lattice,
comparing results of the two-cluster-marginal SDP for various cluster
sizes. The results are shown in Figure \ref{fig:tfi_4x3}. Here we are more limited 
by the preliminary implementation in what can be tested, though the observations 
are compatible with those preceding remarks which are applicable. In Figure \ref{fig:tfi_4x3_noGC}, we once 
again test the effect of removing the global semidefinite constraints, and similar conclusions apply.

\subsection{Anti-ferromagnetic Heisenberg model}
\label{sec:afh}
First we consider the anti-ferromagnetic Heisenberg model (\ref{eq:afh})
on a periodic $12\times1$ lattice, comparing results of the
two-cluster-marginal SDP for various cluster sizes. We also test the
periodicity constraints of section~\ref{sec:periodicityConstraints} and the
case of overlapping clusters, as well as the effect of omitting the 
global semidefinite constraints. The results are shown in Table \ref{table:afh_12x1}.

 \begin{table}
\centering{}\renewcommand{\arraystretch}{1.3}%
\begin{tabular}{|c|c|c|c|c|c|}
\cline{2-6} 
\multicolumn{1}{c|}{}& $1\times1$ & $2\times1$ & $2\times1$, p.c. & $2\times1$, overlap & $3\times1$, p.c.\tabularnewline
\hline 
{With global constraints} & 0.6017 & 0.0634 & 0.0462 & 0.0159 & 0.0048\tabularnewline
\hline 
{Without global constraints} & 1.2042 & 0.2042 & 0.2042 & 0.2042 & 0.0310\tabularnewline
\hline 
\end{tabular}\caption{Energy error by cluster specification for the AFH model on periodic $12\times1$ lattice. For reference, the exact ground state energy is $-1.7958$. 
Approximate energies for the first line are
computed via the two-cluster-marginal relaxation. 
Note that `p.c.'
indicates the inclusion of the periodicity constraints introduced
in section \ref{sec:periodicityConstraints}, and `overlap' indicates
that the choice of overlapping $2\times1$ clusters, i.e., $\{1,2\},\{2,3\},\{3,4\},\ldots,\{11,12\}$. 
For the results of the second line, the global semidefinite constraints were omitted.
}
\label{table:afh_12x1}
\end{table}

 In Table \ref{table:afh_4x3} we show results for the AFH model on
 a periodic $4\times3$ lattice for various cluster sizes. For these 
 experiments, the observations are qualitatively similar to those 
 reported for the TFI model, though the relative energy errors are 
 larger. In particular, the errors for $1\times 1$ clusters are quite large,
 though the error falls dramatically as the cluster size is increased. Moreover,
 the global constraints achieve significant error reduction even though the
 Hamiltonian is local.

\begin{table}
\centering{}\renewcommand{\arraystretch}{1.3}%
\begin{tabular}{|c|c|c|c|}
\cline{2-4} 
\multicolumn{1}{c|}{} & $1\times1$ clusters & $2\times1$ clusters & $1\times3$ clusters\tabularnewline
\hline 
With global constraints & 1.0439 & 0.3937 & 0.0410\tabularnewline
\hline 
 Without global constraints & 3.5439 & 2.1897 & 0.8773\tabularnewline
\hline 
\end{tabular}\caption{Energy error by cluster specification for the AFH model on periodic $4\times3$ lattice. For reference, the exact ground state energy is $-2.4561$. Approximate energies for the first line are
computed via the two-cluster-marginal relaxation. Approximate energies for the second line 
are obtained by omitting the global semidefinite constraints.}
\label{table:afh_4x3}
\end{table}

\FloatBarrier

\subsection{Hubbard model}
\label{sec:hubbard}
\begin{figure}[h]
\begin{centering}
\includegraphics[bb=40bp 0bp 520bp 420bp,scale=0.45]{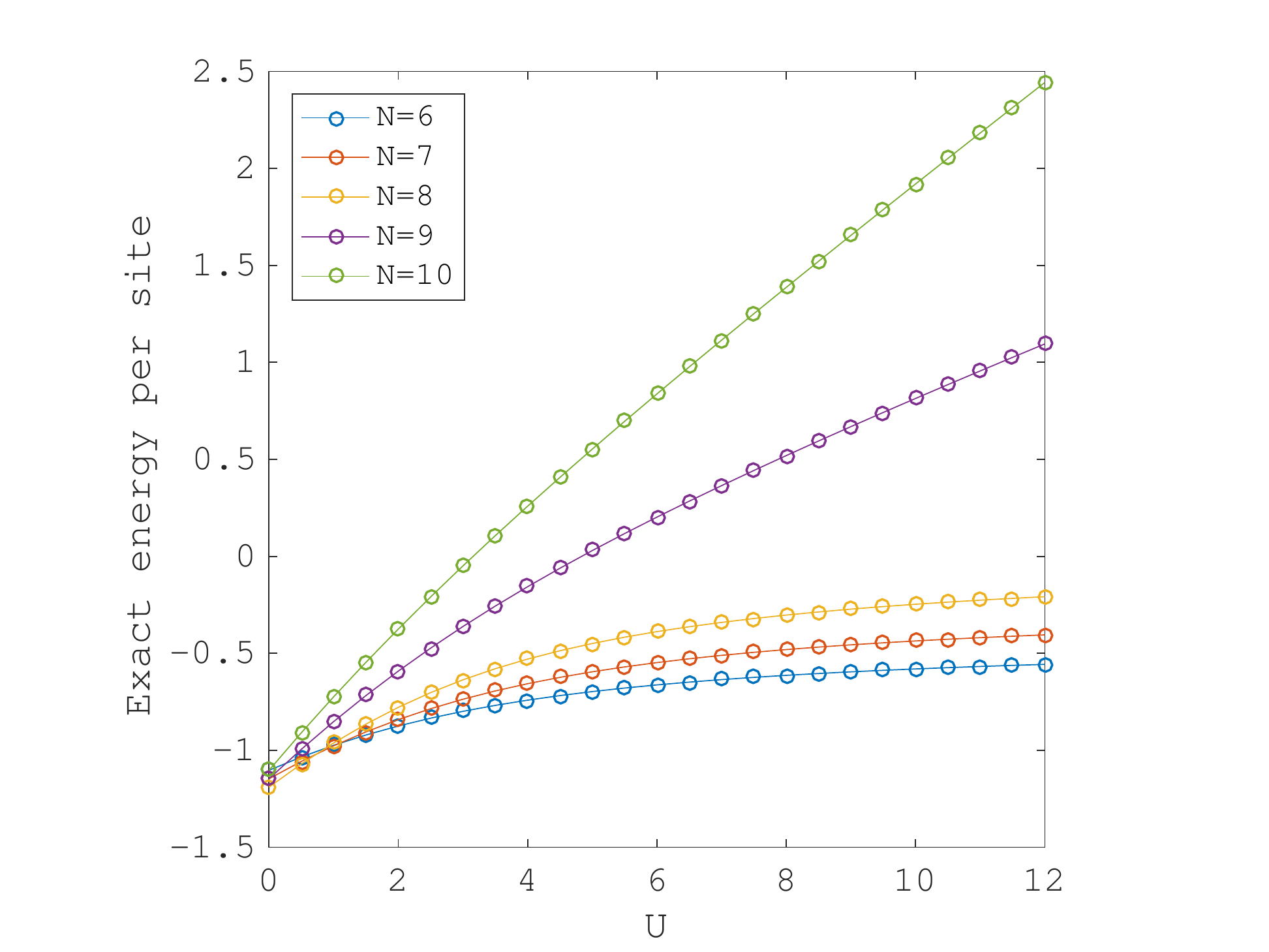}\includegraphics[bb=40bp 0bp 520bp 420bp,scale=0.45]{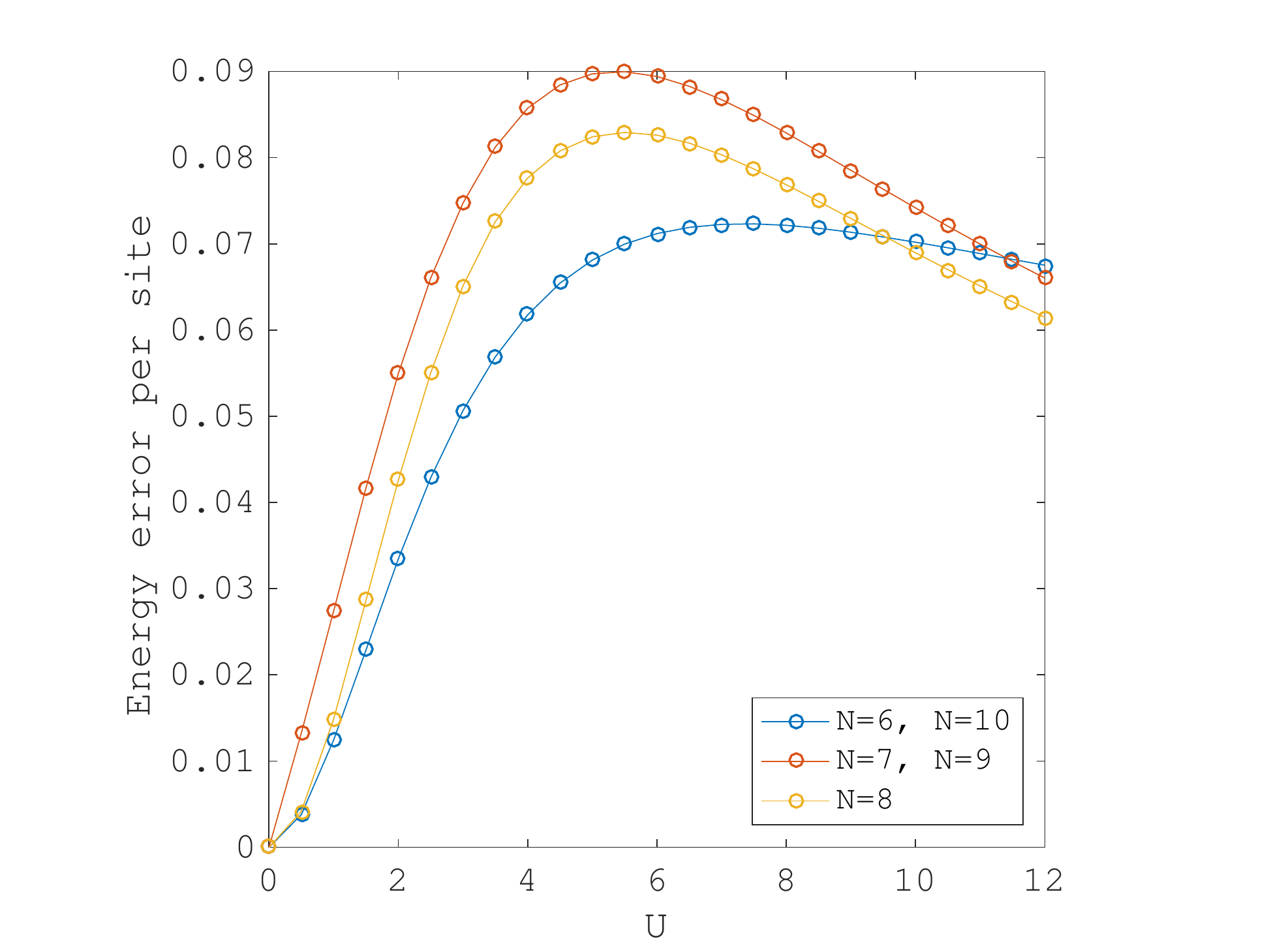}
\par\end{centering}

\caption{Hubbard model on non-periodic $8\times1$ lattice. Approximate energies
are computed via the two-cluster-marginal relaxation with $1\times1$
clusters $C_{i}:=\{(i,\uparrow),(i,\downarrow)\}$. Note that the
energy errors in the cases $N=6$ and $N=7$ coincide with the errors
in the cases $N=10$ and $N=9$, respectively due to the particle-hole symmetry.}
\label{fig:hubbard}
\end{figure}

Finally we consider the Hubbard model (\ref{eq:hubbard}) on a non-periodic
$8\times1$ lattice with particle numbers $N=6,7,8,9,10$ and interaction
strengths $U\in[0,12]$. In Figure \ref{fig:hubbard}, we plot results
for the two-cluster-marginal relaxation with $1\times1$ clusters
$C_{i}:=\{(i,\uparrow),(i,\downarrow)\}$. Observe that for $U=0$,  the system is non-interacting and the energy is exact, 
as guaranteed by the discussion in section \ref{sec:exact_noninteracting}. Furthermore, the error of the energy decreases with respect to
$U$ (even without normalizing by $U$). We remark that the error of the energy per site is on par with that of DMET~\cite{KniziaChan2012} when the same cluster sizes are used. In comparison to DMET, variational embedding is less accurate for intermediate $U$ (i.e., $U\approx 4$) 
but scales more gracefully in the regime of large $U$ (i.e., $U \gtrsim 8$). However, a thorough comparison 
of variational embedding with other embedding methods will be a matter for future work following 
more careful implementation.

\begin{figure}
\centering{}\includegraphics[bb=40bp 0bp 520bp 420bp,scale=0.45]{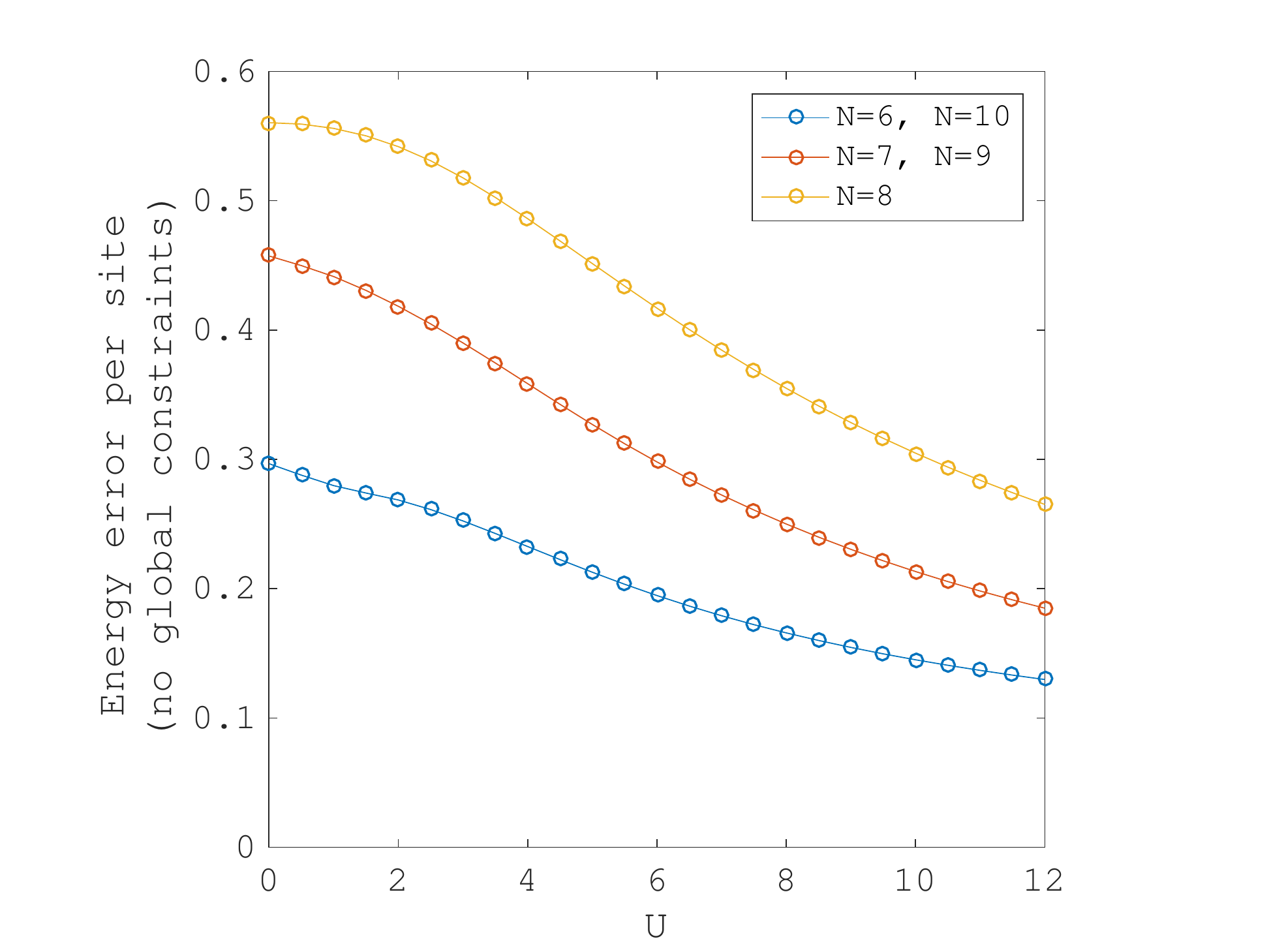}\caption{Results for the same model and same relaxation as in Figure \ref{fig:hubbard}, with the modification that the 
global semidefinite constraints are omitted in all cases. Note the change of scale of the vertical axis relative to the analogous plot of Figure \ref{fig:hubbard}.}
\label{fig:hubbard_noGC}
\end{figure}

In Figure \ref{fig:hubbard_noGC} we test the same relaxation on the same model problems, except that 
once again we \emph{omit the global semidefinite constraints}. Once again we observe significant degradation of the lower bound. Note moreover that the omission of these 
constraints breaks the exactness of the relaxation energy for $U=0$.

\FloatBarrier

\section{Duality and the effective Hamiltonian perspective}
\label{sec:duality}
In order to reduce the computational cost for solving the SDP in the
variational embedding (called the primal problem), we may consider the
associated dual problem.  For simplicity, we consider duality only for the
two-marginal SDP in the quantum spin setting, and it will be convenient to
take the `abstract perspective' of section \ref{sec:abstractGlobal}, with
possibly restricted operator sets as in Remark \ref{rem:restrictedOpSets}.
Duality in other settings can be approached by similar means.

\subsection{The quantum Kantorovich problem}
\label{sec:kantorovich}
In preparation for our discussion of the duality of the two-marginal
SDP, we first introduce the notion of the quantum Kantorovich problem,
which is a direct quantum analog (and in fact generalization) of the
Kantorovich problem of optimal transport~\cite{villani2008optimal}. See 
also 
in~\cite{ChenGangboEtAl2019,GolseEtAl2016,DattaRouze2017,ZhouYingEtAl2019,CagliotiGolsePaul2018}
for related, though different, presentations.\\

The analogy to classical optimal transport is defined by
replacing probability measures with density operators, a cost function
with a cost operator $\hat{C}$, and classical marginalization with
quantum marginalization (i.e., the partial trace). Given operators
$\mu_{i}\in\mathrm{End}(Q_{i})$ for $i=1,2$ of unit trace, we may
define the optimal quantum Kantorovich cost via the SDP 
\begin{eqnarray*}
\mathbf{QK}[\hat{C}\,;\,\mu_{1},\,\mu_{2}]\ :=\ \underset{\pi\in\mathrm{End}(Q_{1}\otimes Q_{2})}{\mbox{minimize}} \quad &  & \Tr[C\pi]\\
\mbox{subject to} \quad &  & \pi\succeq0\\
 &  & \mu_{1}=\Tr_{\{2\}}[\pi],\ \mu_{2}=\Tr_{\{1\}}[\pi].
\end{eqnarray*}
 Note that if $\mu_{1}\not\succeq0$ or $\mu_{2}\not\succeq0$, then
since $\pi\succeq0$ implies that $\Tr_{\{i\}}[\pi]\succeq0$, the
problem is infeasible, i.e., $\mathbf{QK}[C\,;\,\mu_{1},\,\mu_{2}]=+\infty$.
Hence without loss of generality one may assume that $\mu_{i}\succeq0$,
i.e., that the $\mu_{i}$ are indeed density operators on $Q_{i}$.
Nonetheless, the slightly relaxed perspective will be of some use
below. In fact, conversely, the program \emph{is} feasible whenever
$\mu_{1},\mu_{2}\succeq0$ because in this case $\pi=\mu_{1}\otimes\mu_{2}$
is a feasible point.\\

There is a notion of quantum Kantorovich duality that analogizes the
usual notion, as follows. Let the \emph{Hermitian} operators $A\in\mathrm{End}(Q_{1})$
and $B\in\mathrm{End}(Q_{2})$ be dual variables for the first and
second marginal constraints, respectively. These will be the `quantum
Kantorovich potentials.' Dualizing these constraints yields the Lagrangian
\[
\mathcal{L}_{\mathrm{QK}}(\pi,A,B)=\Tr[C\pi]+\Tr[A(\mu_{1}-\Tr_{\{2\}}[\pi])]+\Tr[B(\mu_{2}-\Tr_{\{1\}}[\pi])]
\]
 still constrained by $\pi\succeq0$. Using the fact that $\Tr[A\,\Tr_{\{2\}}[\pi]]=\Tr[(A\otimes\mathrm{Id})\pi]$
and $\Tr[B\,\Tr_{\{1\}}[\pi]]=\Tr[(\mathrm{Id}\otimes B)\pi]$, we
obtain 
\[
\mathcal{L}_{\mathrm{QK}}(\pi,A,B)=\Tr[A\mu_{1}]+\Tr[B\mu_{2}]+\Tr[\left(C-A\otimes\mathrm{Id}-\mathrm{Id}\otimes B\right)\pi].
\]
Now for fixed $A,B$, we have 
\[
\inf_{\pi\succeq0}\Tr[\left(C-A\otimes\mathrm{Id}-\mathrm{Id}\otimes B\right)\pi]=\begin{cases}
0, & C-A\otimes\mathrm{Id}-\mathrm{Id}\otimes B\succeq0\\
-\infty, & \mathrm{otherwise}.
\end{cases}
\]
 Hence we have derived the Kantorovich dual problem 
\begin{eqnarray}
\underset{A,B\ \mathrm{Hermitian}}{\mbox{maximize}}\quad &  & \Tr[A\mu_{1}]+\Tr[B\mu_{2}]\label{eq:Kdual}\\
\mbox{subject to}\quad &  & A\otimes\mathrm{Id}+\mathrm{Id}\otimes B\preceq C.\nonumber 
\end{eqnarray}
Strong duality holds by Sion's minimax theorem~\cite{Komiya1988} (together with the
compactness of the feasible set of the primal problem).\\

Let $\pi$ be the minimizer for the primal problem, and suppose that
the dual problem admits a maximizer $(A,B)$. Then let $M=C-A\otimes\mathrm{Id}-\mathrm{Id}\otimes B$,
so 
\[
\Tr[M\pi]=\Tr[C\pi]-\Tr[(A\otimes\mathrm{Id})\pi]-\Tr[(\mathrm{Id}\otimes B)\pi]=\Tr[C\pi]-\Tr[A\mu_{1}]-\Tr[B\mu_{2}]=0,
\]
 by primal and dual optimality. But $\pi\succeq0$, so we can write
$\pi=\sum_{i=1}^{m}p_{i}\,\phi_{i}\phi_{i}^{*}$ where $p_{i}>0$,
and $\Tr[M\pi]=\sum_{i=1}^{m}p_{i}\,\phi_{i}^{*}M\phi_{i}$. But also
$M\succeq0$, so $p_{i}\:\phi_{i}^{*}M\phi_{i}\geq0$ for all $i=1,\ldots,m$.
Then since $\Tr[M\pi]=0$ it follows that $\phi_{i}^{*}M\phi_{i}=0$
for all $i=1,\ldots,m$, and since $M\succeq0$ this means that $M\phi_{i}=0$
for all $i=1,\ldots,m$.\\

Therefore $\pi$ is a convex combination of orthogonal projectors
onto mutually-orthogonal, degenerate ground state eigenvectors of
the Hamiltonian $C-A\otimes\mathrm{Id}-\mathrm{Id}\otimes B$. For the reader familiar 
with optimal transport, we remark that this
observation generalizes the corresponding observation~\cite{villani2008optimal} in the classical setting
on the support of the Kantorovich coupling, i.e., that $\pi_{ij}\geq0$
only if $\phi_{i}+\psi_{j}=c_{ij}$, where $\pi=(\pi_{ij})$, $\phi=(\phi_{i})$
and $\psi=(\psi_{i})$ are the Kantorovich potentials, and $c=(c_{ij})$
is the cost matrix. \\

In fact, one can consider a regularization of the primal problem by
a von Neumann entropy penalty (scaled by $\beta$), for which the
solution can be shown to be of the form 
\[
\pi_{\beta}=\frac{1}{Z_{\beta}}\exp\left[-\beta(C-A_{\beta}\otimes\mathrm{Id}-\mathrm{Id}\otimes B_{\beta})\right],
\]
 where $A_{\beta}$ and $B_{\beta}$ are the unique operators chosen
to yield the desired marginals $\mu_{1},\mu_{2}$. 
This is the quantum analogy of the entropic regularization of 
classical optimal transport~\cite{Cuturi2013}.
In the `zero-temperature'
limit $\beta\ra\infty$ one expects $\pi_{\beta}\ra\pi$, $A_{\beta}\ra A$,
and $B_{\beta}\ra B$.

\subsection{Partial duality}
\label{sec:partialDual}
Before any derivations, we comment that strong duality (i.e., the fact 
that there is zero gap between the optimal values of the primal and dual problems 
for the two-marginal SDP) can 
be understood as follows. In the original primal problem (\ref{eq:sdpObj}), 
the feasible domain for $\{\rho_{i}\},\,\{\rho_{ij}\}_{i<j}$
in this problem is compact, so strong duality holds simply by Sion's minimax theorem~\cite{Komiya1988}. 
The question of whether the dual optimizer is attained
is more subtle and will be deferred to future work, though see~\cite{KhooEtAl2019}
for the discussion of strong duality in a similar setting.\\

Now we turn to the derivation of the partial dual problem. 
We adopt the `abstract' perspective on the global semidefinite constraints 
introduced in section~\ref{sec:abstractGlobal}, as well as the notation of that section.
Referring to (\ref{eq:sdpObj}),
we first consider a \emph{partial }Lagrangian obtained by dualizing
\emph{only }the constraint (\ref{eq:sdpC4}): 

\[
\mathcal{L}_{\mathrm{part}}\left(\{\rho_{i}\},\{\rho_{ij}\},X\right)=\sum_{i}\Tr\left[H_{i}\rho_{i}\right]+\sum_{i<j}\Tr\left[H_{ij}\rho_{ij}\right]-\Tr\left(G[\{\rho_{ij}\}]\,X\right),
\]
 whose domain is defined by $X\in\mathbb{C}^{\left(\sum_{i}m_{i}^{2}\right)\times\left(\sum_{i}m_{i}^{2}\right)}$
Hermitian positive semidefinite and $\{\rho_{i}\},\{\rho_{ij}\}$
satisfying constraints (\ref{eq:sdpC1}), (\ref{eq:sdpC2}), and (\ref{eq:sdpC3}).

Now 
\begin{eqnarray*}
\Tr\left(G[\{\rho_{ij}\}]\,X\right) & = & \sum_{ij}\Tr\left(G_{ij}[\rho_{ij}]\,X_{ji}\right)\\
 & = & \sum_{i}\sum_{\alpha\beta}\Tr\left[\rho_{i}O_{i,\alpha}^{\dagger}O_{i,\beta}\right](X_{ii})_{\beta\alpha}+\sum_{i\neq j}\sum_{\alpha\beta}\Tr\left[\rho_{ij}\left(O_{i,\alpha}^{\dagger}\otimes O_{j,\beta}\right)\right](X_{ji})_{\beta\alpha}\\
 & = & \sum_{i}\sum_{\alpha\beta}\Tr\left[\rho_{i}O_{i,\alpha}^{\dagger}O_{i,\beta}\right](X_{ii})_{\beta\alpha}\\
 &  & \ \ +\ \sum_{i<j}\sum_{\alpha\beta}\left\{ \Tr\left[\rho_{ij}\left(O_{i,\alpha}^{\dagger}\otimes O_{j,\beta}\right)\right](X_{ji})_{\beta\alpha}+\Tr\left[\rho_{ji}\left(O_{j,\beta}^{\dagger}\otimes O_{i,\alpha}\right)\right](X_{ij})_{\alpha\beta}\right\} .
\end{eqnarray*}
Now by the hermiticity of $X$ we have $(X_{ji})_{\beta\alpha}=\overline{(X_{ij})_{\alpha\beta}}$,
and we also have the identity 
\[
\Tr\left[\rho_{ji}\left(O_{j,\beta}^{\dagger}\otimes O_{i,\alpha}\right)\right]=\Tr\left[\rho_{ij}\left(O_{i,\alpha}\otimes O_{j,\beta}^{\dagger}\right)\right].
\]
 Therefore 
\[
\Tr\left(G[\{\rho_{ij}\}]\,X\right)=\sum_{i}\Tr\left[Y_{i}(X_{ii})\,\rho_{i}\right]+\sum_{i<j}\Tr\left[Y_{ij}(X_{ij})\,\rho_{ij}\right],
\]
where we have defined the functions $Y_{i}:\mathbb{C}^{\vert\mathcal{I}_{i}\vert\times\vert\mathcal{I}_{i}\vert}\ra\mathrm{End}(Q_{i})$ 
and 
$Y_{ij}:\mathbb{C}^{\vert\mathcal{I}_{i}\vert\times\vert\mathcal{I}_{j}\vert}\ra\mathrm{End}(Q_{i}\otimes Q_{j})$
by 
\[
Y_{i}(M)=\sum_{\alpha\beta}\overline{M}_{\alpha\beta}\,O_{i,\alpha}^{\dagger}O_{i,\beta},\quad Y_{ij}(M)=\left[\sum_{\alpha\beta}\overline{M}_{\alpha\beta}\,\left(O_{i,\alpha}^{\dagger}\otimes O_{j,\beta}\right)\right]+\mathrm{h.c.},
\]
 where `h.c.' denotes the Hermitian conjugate. Note that if $M$ is
Hermitian, then $Y_{i}(M)$ is Hermitian as well, hence $Y_{i}(X_{ii})$
and $Y_{ij}(X_{ij})$ are Hermitian operators.\\

By applying Sion's minimax theorem~\cite{Komiya1988}
and then separating the infimum over $\{\rho_{i}\},\{\rho_{ij}\}$
into an outer infimum over $\{\rho_{i}\}$ (subject to constraint
(\ref{eq:sdpC3})) and an inner infimum over $\{\rho_{ij}\}$ (subject
to constraints (\ref{eq:sdpC1}) and (\ref{eq:sdpC2})), we may rewrite
the two-marginal SDP energy as 
\begin{equation}
E_{0}^{(2)}=\sup_{X\succeq0}\ \ \underset{\{\rho_{i}\}\,:\,\Tr[\rho_{i}]=1,\forall i}{\inf}\ \ \mathcal{F}\left(X,\{\rho_{i}\}\right),\label{eq:partialDual}
\end{equation}
 where 
\begin{equation}
\mathcal{F}\left(X,\{\rho_{i}\}\right):=\sum_{i}\Tr\left[(H_{i}-Y_{i}(X_{ii}))\rho_{i}\right]+\sum_{i<j}\mathbf{QK}[H_{ij}-Y_{ij}(X_{ij})\,;\,\rho_{i},\,\rho_{j}].\label{eq:partialDualObj}
\end{equation}
 This is the form of a concave-convex maxmin problem. The effective
domain of the minimization over $\{\rho_{i}\}$ is in fact specified
by the constraints $\Tr[\rho_{i}]=1,\ \rho_{i}\succeq0$ for all $i$,
because if $\rho_{i}\not\succeq0$ for some $i$, then at least one
of the quantum Kantorovich problems in the expression for $\mathcal{F}\left(X,\{\rho_{i}\}\right)$
is infeasible, i.e., of infinite optimal cost. The significance of
this form is that for fixed $X,\{\rho_{i}\}$, the two-marginals $\rho_{ij}$
have been entirely decoupled from one another in the evaluation of
$\mathcal{F}\left(X,\{\rho_{i}\}\right)$. Moreover, for each pair
$i<j$, we see the emergence of the effective Hamiltonians $H_{i}^{\mathrm{eff}}(X_{ii}):=H_{i}-Y_{i}(X_{ii})$
and $H_{ij}^{\mathrm{eff}}(X_{ij}):=H_{ij}-Y_{ij}(X_{ij})$ on $Q_{i}$
and $Q_{i}\otimes Q_{j}$, respectively. Notice that the new contributions
to these effective Hamiltonians are linear combinations of operators
of the form $O_{i,\alpha}^{\dagger}O_{i,\beta}$ and $O_{i,\alpha}^{\dagger}\otimes O_{i,\beta}$,
respectively. Thus we see how our choice of effective operator lists
is reflected in the richness of our class of possible effective Hamiltonians.

\subsection{Computational perspective}
\label{sec:compPartialDual}
From the computational point of view, the partial dual formulation
can be much more efficient to solve than the primal formulation. Although general results guarantee that
the complexity of solving the two-marginal SDP (\ref{eq:sdpObj})
is only polynomial in $M$, direct solution of the primal problem (by, e.g., interior-point
methods) may still scale quite poorly in practice. One might hope
that the complexity should be limited only by $O(M^{3})$ per iteration,
i.e., the cost of diagonalizing a matrix of size proportional to $M$,
since the SDP constraint (\ref{eq:sdpC4}) concerns a matrix of size
proportional to $M$. However, since the semidefinite matrix $G$
is entangled with further equality constraints, the best guarantees
for interior-point methods are far more pessimistic. One can interpret
our discussion of duality thus far as revealing a special structure
of these equality constraints that allows us in principle to design
methods achieving a cost of $O(M^{3})$ per iteration. (We remark
that similar considerations could be expected to achieve a cost of
$O(M)$ per iteration for the quasi-local two-marginal SDP with fixed
$d_{\max}$, as described in Remark~\ref{rem:quasilocal}, though we omit details
for simplicity.)\\

Now we describe how to compute gradients of $\mathcal{F}\left(X,\{\rho_{i}\}\right)$, in order to apply, e.g., 
gradient ascent-descent methods. For fixed $X,\{\rho_{i}\}$, let $(A_{ij}^{\star},B_{ij}^{\star})$
be the unique dual optimizer (assuming that it exists) for the Kantorovich
dual formulation of $\mathbf{QK}[H_{ij}-Y_{ij}(X_{ij})\,;\,\rho_{i},\,\rho_{j}]$.
Then it follows that 

\[
\frac{\partial\mathcal{F}}{\partial\rho_{k}}(X,\{\rho_{i}\})=H_{k}-Y_{k}(X_{kk})+\sum_{j>k}A_{kj}^{\star}+\sum_{i<k}B_{ik}^{\star}
\]
 (Note that if the dual optimizer is not unique, one only gets a supergradient.)
One may take a gradient descent step for $\rho_{k}$ in the direction
of the \emph{traceless part} of $\frac{\partial\mathcal{F}}{\partial\rho_{k}}$,
adjusting the step size if necessary to guarantee that $\rho_{k}\succeq0$.
Moreover, letting $\rho_{ij}^{\star}$ be the primal solution of the
Kantorovich problem indicated by $\mathbf{QK}[H_{ij}-Y_{ij}(X_{ij})\,;\,\rho_{i},\,\rho_{j}]$,
we have 
\[
\frac{\partial\mathcal{F}}{\partial(\overline{X}_{ii})_{\alpha\beta}}(X,\{\rho_{i}\})=-\Tr[O_{i,\alpha}^{\dagger}O_{i,\beta}\,\rho_{i}],\quad\frac{\partial\mathcal{F}}{\partial(\overline{X}_{ij})_{\alpha\beta}}(X,\{\rho_{i}\})=-\Tr\left[\left(O_{i,\alpha}^{\dagger}\otimes O_{j,\beta}\right)\rho_{ij}^{\star}\right].
\]
 (If the primal optimizer is not unique, one only gets a subgradient.)
After taking a gradient ascent step in $X$, one may project onto
the feasible domain $\{X\succeq0\}$ by diagonalizing $X$ and zeroing
all negative eigenvalues.\\

Efficient methods for solving the primal and dual quantum Kantorovich
problems (beyond black-box SDP solvers) will be explored in future
work. In particular, preliminary results indicate promise for a quantum
analog of the classical Sinkhorn scaling algorithm~\cite{Cuturi2013}, for which the
computational cost per iteration is roughly given by the cost of diagonalizing
certain operators on $Q_{i}\otimes Q_{j}$.

\subsection{Full duality}
\label{sec:fullDual}
For completeness we also derive the full dual problem to the original two-marginal SDP. We first introduce dual variables
$\lambda_{i}\in\R$ for the constraints $\Tr[\rho_{i}]=1$ appearing
in the minimization within (\ref{eq:partialDual}), and then exchange the resulting internal 
supremum over $\lambda$ with the infimum over $\{\rho_i\}$
to obtain the problem: 
\begin{eqnarray*}
 &  & \sup_{X\succeq0,\,\lambda}\ \ \underset{\{\rho_{i}\}}{\inf}\ \left\{ \sum_{i}\lambda_{i}(1-\Tr[\rho_{i}])+\mathcal{F}\left(X,\{\rho_{i}\}\right)\right\} \\
 & = & \sup_{X\succeq0,\,\lambda}\ \left\{ \sum_{i}\lambda_{i}+\underset{\{\rho_{i}\}}{\inf}\ \left\{ \sum_{i}\Tr\left[(H_{i}-Y_{i}(X_{ii})-\lambda_{i})\rho_{i}\right]+\sum_{i<j}\mathbf{QK}[H_{ij}-Y_{ij}(X_{ij})\,;\,\rho_{i},\,\rho_{j}]\right\} \right\} .
\end{eqnarray*}
Now by substituting the Kantorovich dual expression (\ref{eq:Kdual})
for $\mathbf{QK}$ and then exchanging maximization and minimization,
we obtain the problem:

\begin{eqnarray*}
\underset{X\succeq0,\,\lambda\in\R^{M},\,\{A_{ij}\},\,\{B_{ij}\}}{\mbox{maximize}} \quad &  & \sum_{i}\lambda_{i}+\underset{\{\rho_{i}\}}{\inf}\ \left\{ \sum_{i}\Tr\left[(H_{i}-Y_{i}(X_{ii})-\lambda_{i})\rho_{i}\right]+\sum_{i<j}\Tr\left[A_{ij}\rho_{i}\right]+\sum_{i<j}\Tr\left[B_{ij}\rho_{j}\right]\right\} \\
\mbox{subject to} \quad &  & A_{ij}\otimes\mathrm{Id}+\mathrm{Id}\otimes B_{ij}\preceq H_{ij}-Y_{ij}(X_{ij}),\quad i<j,\\
 &  & X\succeq0.
\end{eqnarray*}
 Now the expression within the infimum in the objective function can
be rewritten 
\[
\sum_{i}\Tr\left[\left(H_{i}-Y_{i}(X_{ii})-\lambda_{i}+\sum_{j>i}A_{ij}+\sum_{j<i}B_{ji}\right)\rho_{i}\right],
\]
 so carrying out the infimum within the objective function, we arrive
at the full dual: 
\begin{eqnarray*}
\underset{X\succeq0,\,\lambda\in\R^{M},\,\{A_{ij}\},\,\{B_{ij}\}}{\mbox{maximize}} \quad &  & \mathbf{1}^{\top}\lambda\\
\mbox{subject to} \quad &  & H_{i}-Y_{i}(X_{ii})-\lambda_{i}+\sum_{j>i}A_{ij}+\sum_{j<i}B_{ji}=0,\quad i=1,\ldots,M,\\
 &  & A_{ij}\otimes\mathrm{Id}+\mathrm{Id}\otimes B_{ij}\preceq H_{ij}-Y_{ij}(X_{ij}),\quad1\leq i<j\leq M,\\
 &  & X\succeq0,
\end{eqnarray*}
 where the optimization variables $A_{ij}\in\mathrm{End}(Q_{i})$
and $B_{ij}\in\mathrm{End}(Q_{j})$ are understood to be Hermitian.

\bibliography{varembed}
\bibliographystyle{siam}

\end{document}